%% file: main.tex
\documentclass{article}
\usepackage[utf8]{inputenc}
\usepackage{subfig}
\usepackage{tikz}  
\usetikzlibrary{shapes,arrows}

\usepackage[graphicx]{realboxes}
\usepackage{adjustbox}
\usepackage{multirow,array}
\usepackage{floatrow}
\usepackage{natbib}

\newfloatcommand{capbtabbox}{table}[][\FBwidth]
\newcolumntype{P}[1]{>{\centering\arraybackslash}p{#1}}

\usepackage{blindtext}

\usepackage[utf8]{inputenc} 
\usepackage[T1]{fontenc}    
\usepackage{hyperref}       
\usepackage{url}            
\usepackage{booktabs}       
\usepackage{amsfonts}       
\usepackage{nicefrac}       
\usepackage{microtype}      
\usepackage{graphicx}
\usepackage{amsmath}
\usepackage{booktabs}
\usepackage{verbatim}
\usepackage{mathtools}

\usepackage{color}
\newcommand{\todo}[1]{{\color{red} #1 }}
\newcommand{\method}{{\texttt{next-gen-scraPy}}}
\newcommand{\Int}{\int\limits}

\begin{document}

\title{{\method}:  \\Extracting NFL Tracking Data from Images to Evaluate Quarterbacks and Pass Defenses}
\author{Sarah Mallepalle$^1$, Ronald Yurko$^1$, \\Konstantinos Pelechrinis$^2$, Samuel L. Ventura$^1$}
\date{%
    \footnotesize
    $^1$Department of Statistics \& Data Science, Carnegie Mellon University\\%
    $^2$School of Computing and Information, University of Pittsburgh \\
    \today
}

\maketitle



\begin{abstract}

The NFL collects detailed tracking data capturing the location of all players and the ball during each play. Although the raw form of this data is not publicly available, the NFL releases a set of aggregated statistics via their Next Gen Stats (NGS) platform.  They also provide charts showing the locations of pass attempts and outcomes for individual quarterbacks. Our work aims to partially close the gap between what data is available privately (to NFL teams) and publicly, and our contribution is twofold.  First, we introduce an image processing tool designed specifically for extracting the raw data from the NGS pass charts.  We extract the pass outcome, coordinates, and other metadata.  Second, we analyze the resulting dataset, examining the spatial tendencies and performances of individual quarterbacks and defenses.  We use a generalized additive model for completion percentages by field location.  We introduce a Naive Bayes approach for estimating the 2-D completion percentage surfaces of individual teams and quarterbacks, and we provide a one-number summary, completion percentage above expectation (CPAE), for evaluating quarterbacks and team defenses.  We find that our pass location data closely matches the NFL's tracking data, and that our CPAE metric closely matches the NFL's proprietary CPAE metric.
\end{abstract}


\section{Introduction}
\label{sec:intro}
\input{1_Introduction.tex}

\tikzstyle{block} = [rectangle, draw, fill=blue!20, 
    text width=5em, text centered, rounded corners, minimum height=4em]
\tikzstyle{blockred} = [rectangle, draw, fill=red!20, 
    text width=5em, text centered, rounded corners, minimum height=4em]
\tikzstyle{line} = [draw, -latex']

\begin{figure}
\begin{tikzpicture}[node distance = 3cm, auto]

    \node [block] (b1) {Scraping NGS HTML};
    \node [block, right of=b1] (b2) {Collecting Raw Images};
    \node [block, right of=b2] (b3) {Image Preprocessing};
    
    \node [block, below of=b3] (b4) {Cropping Field};
    \node [block, left of=b4] (b5) {Undistortion of Field};
    \node [block, left of=b5] (b6) {Removal of Extraneous Detail};
    
    \node [block, below of=b6] (b7) {Color Thresholding};
    \node [block, right of=b7] (b8) {Pass Extraction with Clustering};
    \node [block, right of=b8] (b9) {Map Pixels to Field Locations};
    
    \node [blockred, below of=b9] (b10) {Output Data};
    
    \path [line] (b1) -- (b2);
    \path [line] (b2) -- (b3);
    \path [line] (b3) -- (b4);
    
    \path [line] (b4) -- (b5);
    
    \path [line] (b5) -- (b6);
    \path [line] (b6) -- (b7);
   \path [line] (b7) -- (b8);
    \path [line] (b8) -- (b9);
    
    \path [line] (b9) -- (b10);
    
\end{tikzpicture}
\caption{Flow chart of the {\method} system} 
\label{fig:flow}
\end{figure}
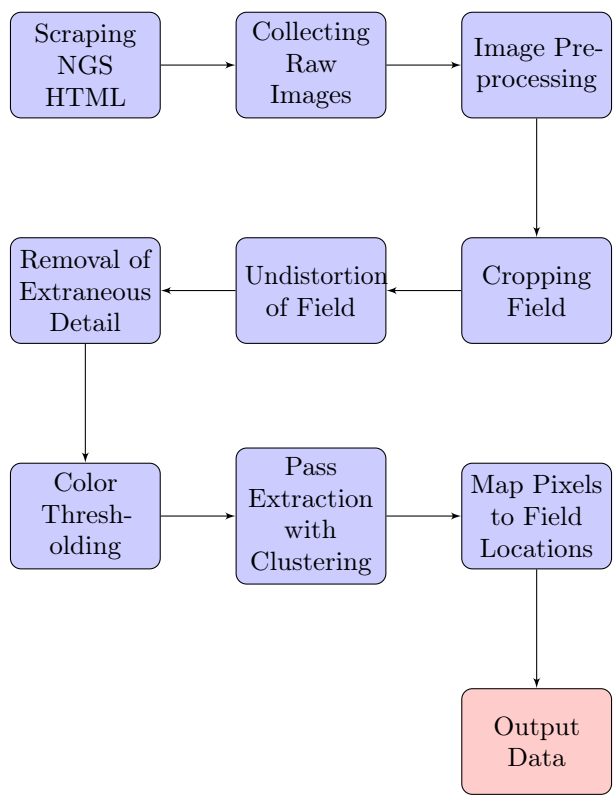


\section{Image Processing Methods for Pass Location Data Extraction}
\label{sec:ngs-scrapy}
\input{2_Methods.tex}

\section{Evaluating and Characterizing Passers and Pass Defenses}
\label{sec:analysis}
\input{3_Analysis.tex}

\section{Results}
\label{sec:results}
\input{4_Results.tex}

\section{Discussion and Conclusions}
\label{sec:conclusions}

\input{5_Discussion.tex}


\bibliographystyle{DeGruyter}
\bibliography{bibliography}

\end{document}

%% file: 1_Introduction.tex
Player tracking data captures the position and trajectory of all athletes and objects of interest (e.g. balls, pucks, etc) on the playing surface for a given sport.  The importance of this data in analyzing the performances and strategies of players and teams has risen dramatically over the past decade, as organizations look to gain an edge over their opponents in ways that were previously not possible.  Publicly, analysis of player tracking data across the four major sports has also increased, but is limited by the availability of such datasets to the public.  Our work aims to bridge the gap between public and private data availability, and to provide an analysis of individual and team passing tendencies in the National Football League (NFL).

\subsection{Player Tracking Data in Professional Sports}

Major League Baseball (MLB) has been tracking pitch trajectory, location, and speed since 2006 with PITCHf/x \citep{what-is-pitchfx}. In 2015, MLB launched Statcast, which additionally tracks the exit velocity and launch angle of a batted ball along with location and movements of every player during a game \citep{statcast}. 
The National Basketball Association (NBA) mandated the installation of an optical tracking system in all stadiums in the 2013-14 season \citep{nba-sportvu}.
This system captures the location of all the players on the court and the ball at a rate of 25 times per second (25 Hz).
This data is further annotated with other information, such as event tracking (``play-by-play''), current score, shot clock, time remaining, etc. 
The National Hockey League (NHL) plans to begin the league-wide use of player- and puck-tracking technology in the 2019-20 season \citep{nhl-tracking}.
The NFL installed a player tracking system in all of its venues during the 2015 season \citep{nfl-tracking}.
NFL's tracking system is RFID-based and records the location of the players and the ball at a frequency of 12.5 Hz. 

This type of data have spurred innovation by driving a variety of applications. 
For example, \cite{cervone2016nba} computed the basketball court's Voronoi diagram based on the players' locations and formalized an optimization problem that provides court realty values for different areas of the court. 
This further allowed the authors to develop new metrics for quantifying the spacing and the positioning of a lineup/team. 
As another example, {\em ghosting} models have been developed in basketball and soccer when tracking data is available  \citep{Le2017CoordinatedMI, grandland}. 
The objective of these models is to analyze the players' movements and identify the optimal locations for the defenders, and consequently evaluate their defensive performance. 
Other models driven by player spatio-temporal data track the possible outcomes of a possession as it is executed, allowing to evaluate a variety of (offensive and defensive) actions that can contribute to scoring but are not captured in traditional boxscore statistics \citep{cervone2016multiresolution,fernandezdecomposing}, while \cite{seidlbhostgusters} further used optical tracking data to learn how a defense is likely to react to a specific offensive set in basketball using reinforcement learning. 
Recently, \cite{burkedeepqb} developed a model using player tracking data from the NFL to predict the targeted receiver, pass outcome and gained yards. 
Coming to soccer, \cite{Power:2017:PCE:3097983.3098051} define and use a supervised learning approach for the risk and reward of a specific pass that can further quantify offensive and defensive skills of players and teams. 
For a complete review of sports research with player tracking data, see \citep{gudmundsson2017spatio}.
 
However, one common trend for player tracking datasets across all sports leagues is its limited availability to the public. 
For example, there are very limited samples of player tracking data for the NBA, and those that exist are mostly from the early days of their player-tracking systems\footnote{\url{https://github.com/linouk23/NBA-Player-Movements}}.  Additionally, while pitch-level data is publicly available from the MLB system\footnote{\url{http://gd2.mlb.com/components/game/mlb}}, the Statcast player and ball location data is not.  

\subsection{Player Tracking Data in the NFL}

In December 2018, the NFL became the first professional sports league to publicly release a substantial subset of its player-tracking data of entire games for its most recent completed season, for two league-run data analysis competitions:  (1) the NFL Punt Analytics competition\footnote{\url{https://www.kaggle.com/c/NFL-Punt-Analytics-Competition}}, and (2) the Big Data Bowl\footnote{\url{https://operations.nfl.com/the-game/big-data-bowl}}. 
In total, tracking data from the punt plays from the 2016 and 2017 seasons, as well as all tracking data from the first six weeks of the 2017 regular season were temporarily released to the public for the purposes of these competitions.  The Big Data Bowl player tracking data was removed shortly after the completion of this competition.  

While this is a tremendous step forward for quantitative football research, the limited scope of the released data limits the conclusions that analysts can draw about player and team performances from the most recent NFL season.  
The only metrics available to fans and analysts are those provided by the league itself through their Next Gen Stats (NGS) online platform.  
The NFL's NGS group uses the league's tracking data to develop new metrics and present aggregate statistics to the fans. 
For example, the NGS website presents metrics such as time-to-throw for a QB, completion probability for a pass, passing location charts, and other metrics. 
However, NGS only provides summaries publicly, while the raw data is not available to analysts or fans.  Additionally, any metrics derived from the player tracking data temporarily made available via the NFL's Big Data Bowl is limited in scope and potentially outdated, as this data only covers the first six weeks of the 2017 season.\footnote{\url{https://twitter.com/StatsbyLopez/status/1133729878933725184}}

\subsection{Our Contributions}

The objective of our work is twofold: 

\begin{enumerate}
    \item We create open-source image processing software, {\method}, designed specifically for extracting the underlying data (on-field location of pass attempt relative to the line of scrimmage, pass outcome, and other metadata) from the NGS pass chart images for regular season and postseason pass attempts from the 2017 and 2018 NFL seasons.
    \item We analyze the resulting dataset, obtaining a detailed view of league-wide passing tendencies and the spatial performance of individual quarterbacks and defensive units.  We use a generalized additive model for modeling quarterback completion percentages by location, and an empirical Bayesian approach for estimating the 2-D completion percentage surfaces of individual teams and quarterbacks.  We use the results of these models to create one-number summary of passing performance, called Completion Percentage Above Expectation (CPAE).  We provide a ranking of QBs and team pass defenses according to this metric, and we compare our version of CPAE with the NFL's own CPAE metric.
\end{enumerate}

Our work follows in the footsteps of \texttt{openWAR} \citep{Baumer15}, \texttt{pitchRx}
\citep{pitchRx}, \texttt{nflscrapR} \citep{nflscrapR}, \texttt{nhlscrapR} \citep{nhlscrapr},  \texttt{Lahman} \citep{Lahman}, \texttt{ballr} \citep{ballr}, and \texttt{ncaahoopR} \citep{ncaahoopR}, who each promote reproducible sports research by providing open-source software for the collection and processing of data in sports.

With {\method}, we rely on a variety of image processing and unsupervised learning techniques. 
The input to {\method} is a pass chart obtained from NFL's NGS, similar to the example in Figure \ref{fig:foles} (which we detail in the following sections). 
The output includes the $(x,y)$ coordinates (relative to the line of scrimmage) for the endpoint of each pass (e.g. the point at which the ball is caught or hits the ground) present in the input image, as well as additional metadata such as the game, the opponent, the result of each pass, etc. 
We then process these data and build a variety of models for evaluating passing performance. 
In particular, we develop spatial models for the target location of the passes at the league, team defense, and individual quarterback (QB) levels.
We use generalized additive models (GAMs) and a 2-D empirical Bayesian approach to estimate completion percentage surfaces (i.e., smoothed surfaces that capture the completion percentage expected in a given location) for individual QBs, team offenses, and team defenses.

The rest of this paper is organized as follows: 
Section \ref{sec:ngs-scrapy} describes the data collected and the processing performed by \method. 
Section \ref{sec:analysis} presents the methods used to analyze the raw pass data obtained from {\method}. 
Section \ref{sec:results} demonstrates the accuracy of the image processing procedure and presents the results of our analyses of individual QBs and team defenses.
Finally, Section \ref{sec:conclusions} concludes our study, presenting future directions and current limitations of {\method}.

%% file: 2_Methods.tex
In this section, we describe the raw data obtained from NGS, and the processing performed by \method. 
We further present the output data provided by our software. 

\subsection{Collecting the Raw Image Data}

\begin{figure}[h!]
  \centering
  \includegraphics[width=0.7\textwidth]{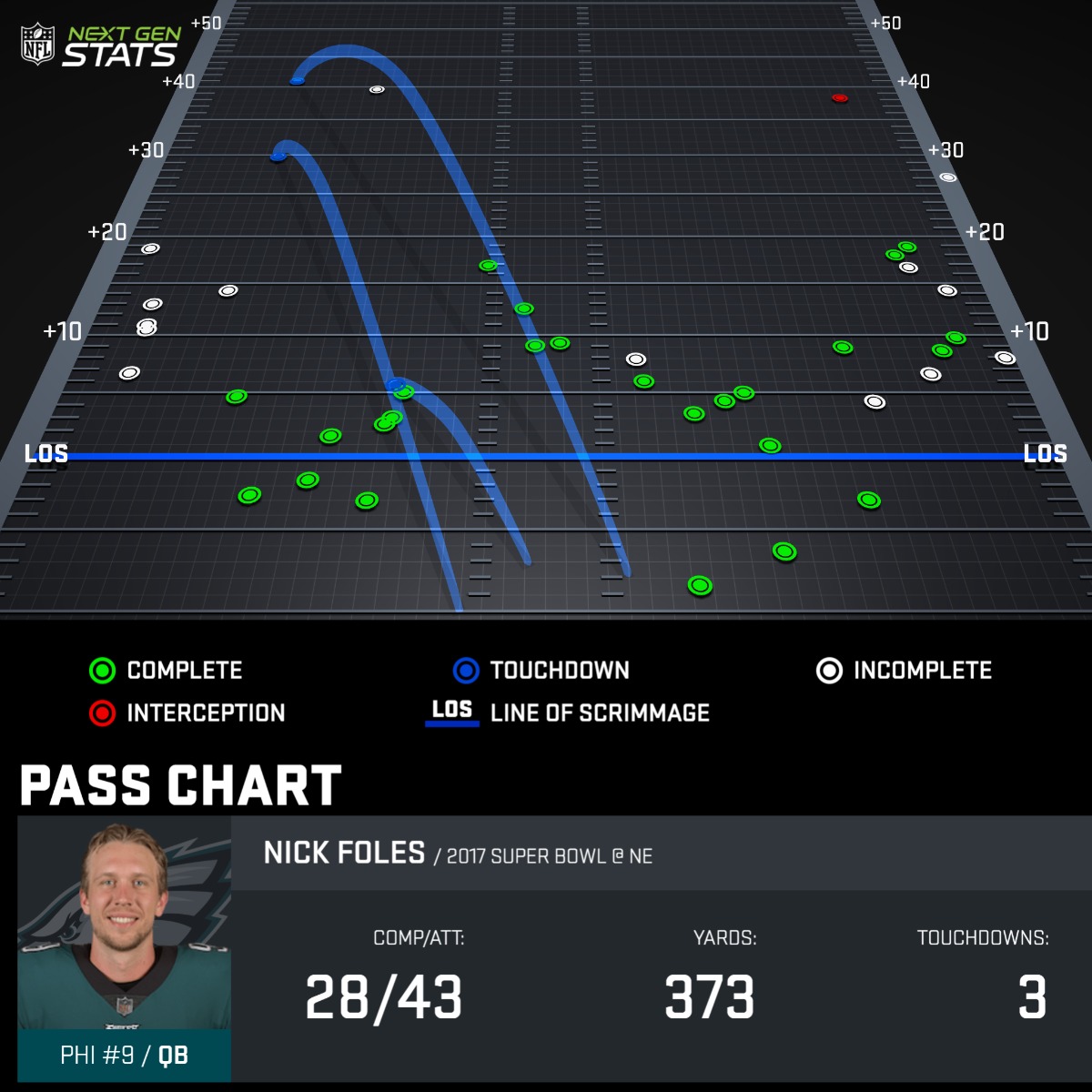}
  \caption{Nick Foles' Pass Chart from Super Bowl LII - Extra-large (1200$\times$1200) image extracted from the HTML of the Next Gen Stats website, which visualizes the location relative to the line of scrimmage of all complete, incomplete, touchdown, and intercepted passes.}
  \label{fig:foles}
\end{figure}

The NFL provides passing charts via their online NGS platform, in the form of JPEG images.  Each chart displays the on-field locations (relative to the line of scrimmage) of every pass thrown by a single quarterback in a single game.  The points, which represent the ending location of each pass (e.g. the point at which a ball is caught or hits the ground) are colored by the outcome of the pass (completion, incompletion, interception, or touchdown).  Each chart is accompanied by a JavaScript Object Notation (JSON) data structure that details metadata about the quarterback or game represented in that chart.  We link each passing chart to the data provided by \texttt{nflscrapR} to obtain additional metadata for each game \citep{nflscrapR}.  These charts are available for most games from the 2017 and 2018 NFL regular and post-seasons, though some are missing from the NGS website without explanation.  An example passing chart is provided in Figure \ref{fig:foles}.

In total, there are 402 pass charts for 248 games throughout the 2017 regular and postseason, and 438 pass charts for 253 games throughout the 2018 regular and postseason. Each pass chart image is of 1200 $\times$ 1200 pixels, and is annotated with metadata that includes information such as player name, team, number of total pass attempts, etc. Appendix \ref{app:appendix-A} provides a detailed list of the metadata necessary to perform data collection and pass detection.

In our analysis, we do not include data from the 2016 season (the first season in which NGS provided these charts), since approximately 65\% of the pass charts are missing. 
Furthermore, the missing charts are not uniformly distributed across teams, biasing potential analyses. 
Finally, the existing 2016 charts frequently do not have the necessary metadata, introducing additional challenges to the data extraction process that we outline below.

\subsection{Image Pre-Processing}

Before being able to extract the raw data from the pass charts, the images must be pre-processed. 
First, the field in the raw image is presented as a trapezoid, rather than a rectangle, which would distort the pass locations extracted from the chart.
Second, the images include unnecessary information that needs to be eliminated in order to allow for the streamlined and accurate processing of pass locations. 

\textbf{Removing Distortion.}  The football field and touchdown pass trajectories as shown in Figure \ref{fig:foles} are distorted, because the trapezoidal projection of the field results in a warped representation of the on-field space, such that (for example) a uniform window of pixels contains more square yards towards the top of the image as compared to the bottom of the image. To fix this, we start by cropping each image to contain only the field, while at the same time we extend the sidelines by 10 yards behind the line of scrimmage, as consistent with every pass chart given by NGS.  
We then project the trapezoidal plane into a rectangular plane that accurately represents the geometric space of a football field. 
In particular, we calculate the homography matrix \textbf{H}, such that every point between each point on the trapezoidal plane $(x_t, y_t)$ is mapped to the rectangular plane $(x_r, y_r)$ through the following equation: 

$$
\begin{bmatrix} 
x_r \\
y_r \\
1
\end{bmatrix} = 
\begin{bmatrix} 
h_{00} & h_{01} & h_{02} \\
h_{10} & h_{11} & h_{12} \\
h_{20} & h_{21} & 1
\end{bmatrix}
\begin{bmatrix} 
x_t \\
y_t \\
1
\end{bmatrix} = \textbf{H} 
\begin{bmatrix} 
x_t \\
y_t \\
1
\end{bmatrix}, 
$$


\noindent where $h_{00}, ... , h_{21}$ are the elements of the homography matrix, \textbf{H}, to obtain the relation between all initial $(x_t, y_t)$ to the corresponding resulting $(x_r, y_r)$ coordinates \citep{szeliski2010computer}. This results in a fully rectangular birds-eye view of the football field, depicting uniform yardage across the height and width of the field.

\begin{figure}[!ht]
   \centering
    \includegraphics[width=.85\textwidth]{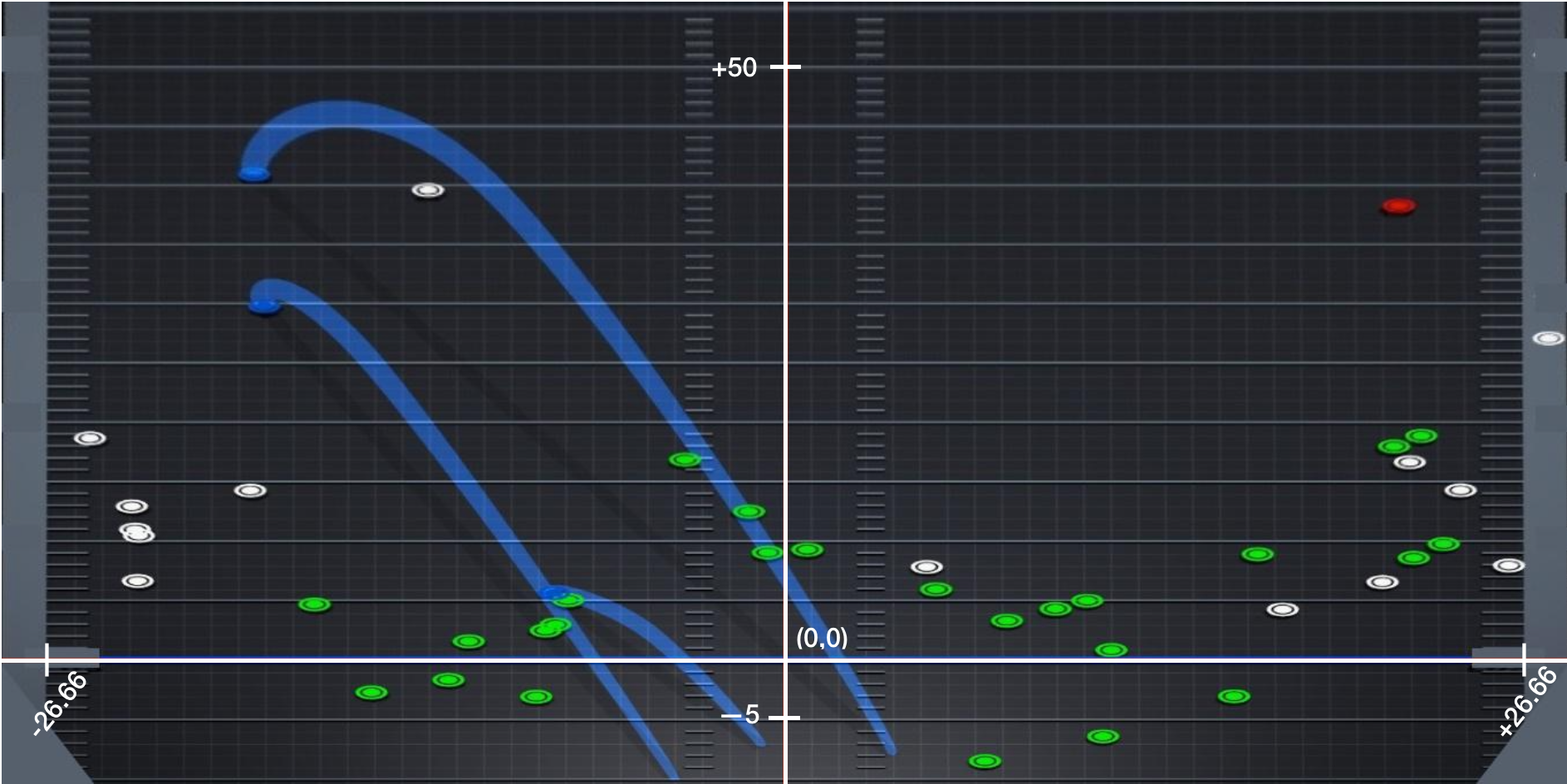}
    \caption{Nick Foles' Pass Chart from Super Bowl LII after fixing the warped perspective of the field and removing the sidelines. The axes  are displayed to show the x-axis falling directly on the line of scrimmage and the y-axis dividing the field vertically in half down the middle.}
   \label{fig:clean}
\end{figure}

\textbf{Eliminating Unnecessary Details from Image.}  
Next, we remove the white sideline numbers of the newly projected field. 
Each pass chart depicts 10 yards behind the line of scrimmage, and also either 55 or 75 yards beyond the line of scrimmage. Based on the pixel height of the newly transformed rectangular field image, we deduce whether or not the field depicts 55 yards (6 sideline markings) or 75 yards (8 sideline markings), and thus find the approximate locations of the white pixels. We then replace the white pixels with the same shade of grey as the sidelines. Removal of the white sideline numbers allows us to easily use a simple white color threshold to extract the incomplete pass locations, as described in Section \ref{sec:clustering}. 
Based on the pixel height and width of the newly transformed rectangular field image, we can calculate the locations of the sideline markings by pixel ratios, and can locate the positions of the sideline markings to remove. 
The final image of the unwarped, clean, field from which the raw locations of the passes are extracted is shown in Figure \ref{fig:clean}(a).

\subsection{Clustering Methods to Extract Pass Locations}

Every pass chart shows the locations of four different types of passes, relative to the line of scrimmage:  completions (green), incompletions (white), touchdowns (blue), and interceptions (red). 
The JSON metadata includes the number of each pass type regardless of whether the pass is being depicted on the image or not\footnote{For example, passes that were thrown out-of-bounds are not depicted in pass charts.}. 
There are a number of technical challenges to overcome when extracting pass locations:

\begin{itemize}
    \item two or more pass locations can overlap,
    \item the number of passes shown on the image does not always match the number of passes given in the JSON metadata of the image,
    \item for touchdown passes, the color of the pass location is the same color as the line of scrimmage and pass trajectory path, rendering color thresholding used for the other types of passes ineffective.
\end{itemize}
To address these issues, \method$~$ combines density-based and distance-based clustering methods (DBSCAN and $K$-means, respectively) with basic image processing techniques to overcome the aforementioned challenges and extract all the pass locations presented on an NGS pass chart. 


\subsubsection{Image Segmentation by Pass Type}
\label{sec:clustering}

\begin{figure}
  \includegraphics[width=.47\textwidth]{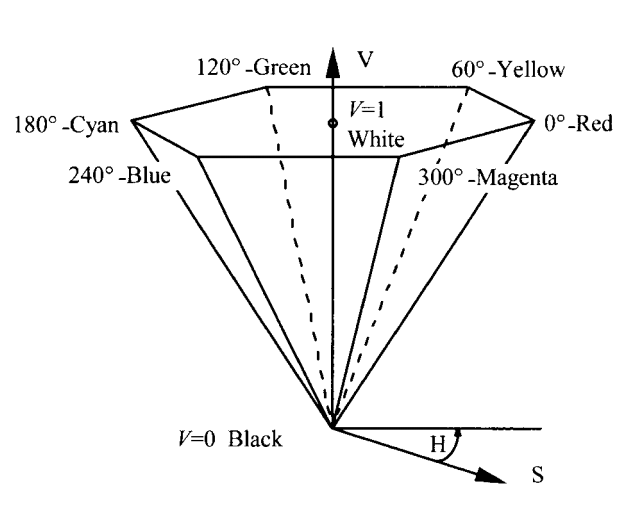}
  \caption{HSV color space representation}
  \label{fig:hsv-figure}
\end{figure}

All four different pass types are marked on the pass chart with different colors. 
Therefore, we examine the Hue-Saturation-Value (HSV) pixel coordinates from the image to identify parts of the image that fall within a specified HSV color range.
In brief, hue characterizes the dominant color contained in the pixel. 
It is captured by the angular position on the color wheel, with red being the reference color (i.e., H = 0 and H = 360). 
Complementary colors are located across of each other on the color wheel and hence, they are 180 degrees apart. 
Saturation measures the color purity, and is captured by the distance of the color from the center of the color wheel.  Value characterizes the brightness of the color \citep{hsv}.  This color system is visualized in Figure \ref{fig:hsv-figure}.

Based on the color of the pass type we want to detect, we use a basic thresholding technique to obtain images where all pixels are black except the ones with the pass locations. 
For example, for completed passes. we will keep the value of pixels within the respective range presented in Table \ref{tab:hsv-table} unchanged, and set the value of all other pixels to 0 corresponding to the black color.

\begin{table}
\begin{center}
 \begin{tabular}{P{20mm}|P{20mm}|P{20mm}} \toprule
  Pass Type & Lower Threshold & Upper  Threshold \\ \midrule
  Complete & (80,100,100) & (160,255,255) \\
  Incomplete & (0,0,90) & (0,0,100) \\ 
  Touchdown & (220,40,40) & (260,100,100) \\ 
  Interception & (0, 60, 60) & (20,100,100) \\ 
\end{tabular}
\caption{HSV color thresholds for every pass type}%
\label{tab:hsv-table}
\end{center}
\end{table}

\begin{figure}[!ht]
   \centering
   \subfloat[][Green thresholding for completions]{\includegraphics[width=.47\textwidth]{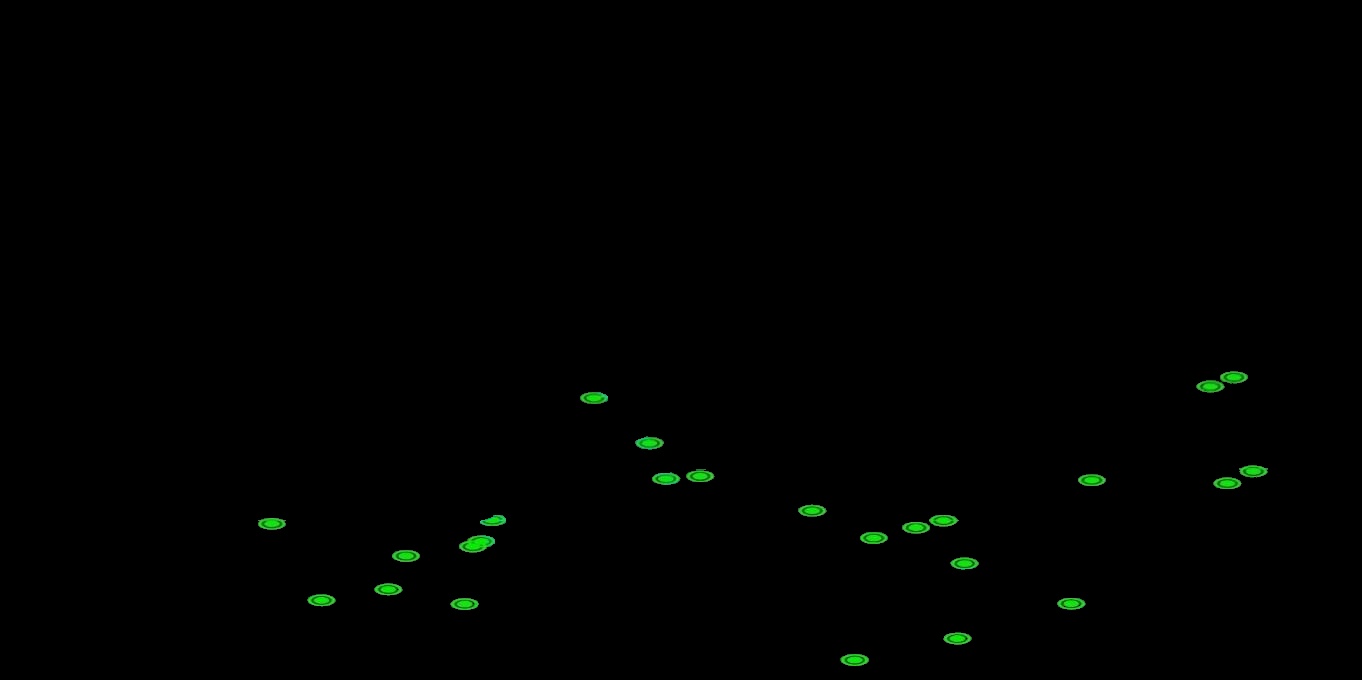}}\quad
   \subfloat[][White thresholding for incompletions]{\includegraphics[width=.47\textwidth]{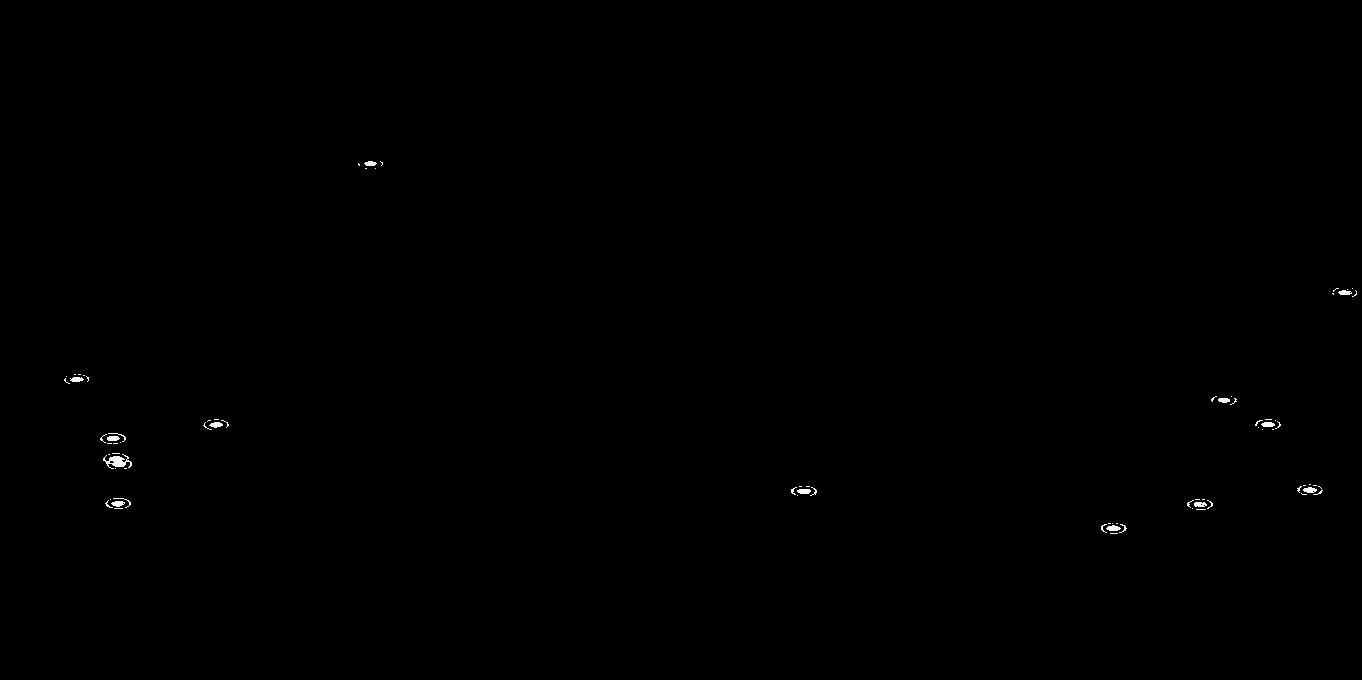}}\\
   \subfloat[][Red thresholding for interceptions]{\includegraphics[width=.47\textwidth]{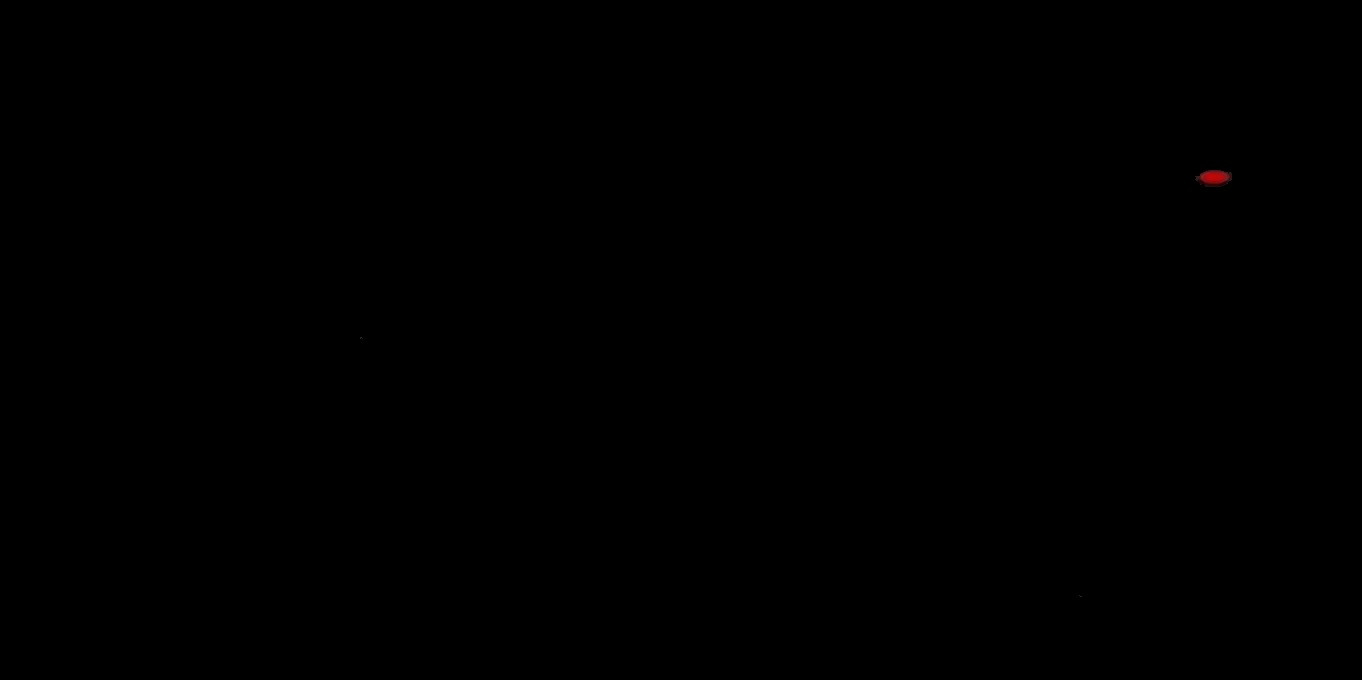}}\quad
   \subfloat[][Blue thresholding for touchdowns (with noise)]{\includegraphics[width=.47\textwidth]{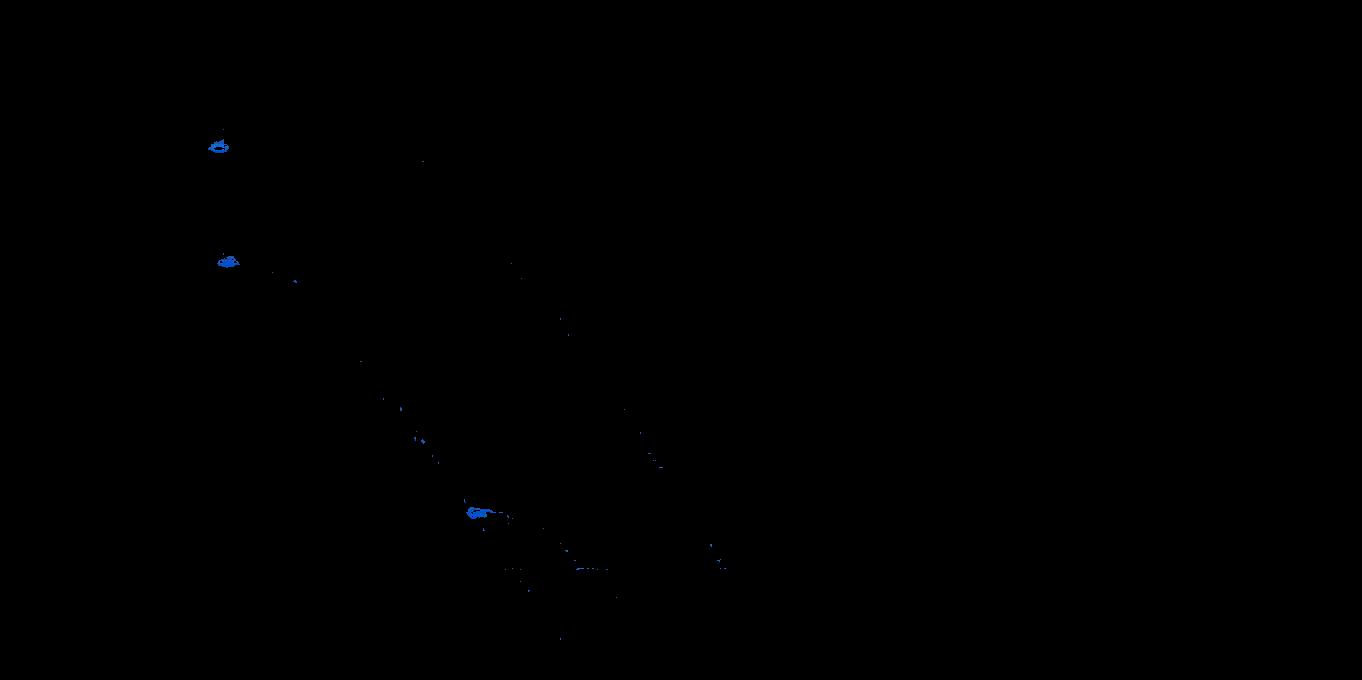}}
   \caption{Color thresholding for all pass types. $K$-means and DBSCAN are subsequently performed on each of these images for pass detection. Noise in the segmented touchdown image results from the pass location having similar color to the line of scrimmage and pass trajectory lines.}
   \label{fig:segmented}
\end{figure}

The resulting segmented image is shown in Figure \ref{fig:segmented}(a), along with the images obtained for the other pass types. As aforementioned in the technical challenges, for the touchdown passes the color thresholding includes additional noise in the final image, as shown in Figure \ref{fig:segmented}(d).


From the JSON metadata associated with the image, we know the number of pass attempts, $n_a$, touchdowns, $n_{td}$, total completions, $n_{tot-c}$, and interceptions, $n_{int}$. The number of touchdowns and interceptions on each image is then simply $n_{td}$ and $n_{int}$, respectively. 
The number of green non-touchdown completions $n_c$ presented on the image excludes the touchdown passes and hence, $n_c = n_{tot-c} - n_{td}$, while the number of gray incompletions is $n_{inc} = n_{a} - n_c - n_{td} - n_{int}$.


Using the segmented images and the number of passes of each type, we detect the pixel locations of each pass type within the respective image (see Sections \ref{sec:kmeans} and \ref{sec:dbscan}). 
We then map these locations to the dimensions of a real football field to identify the on-field locations of each pass in $(x, y)$ coordinates, on the coordinate system presented in Figure \ref{fig:clean}. 
The $y=0$ vertical line runs through the center of the field, while the $x=0$ horizontal line always represents the line of scrimmage.

\subsubsection{Identifying Pass Types with $K$-Means++}
\label{sec:kmeans}



$K$-means is a distance-based clustering method used to define a clustering partition $\mathcal{C}$ of $n$ observations $X_i \in \mathbb{R}^p$, for $i = 1, \hdots, n$, into a pre-determined number of clusters $K$ such that the within-cluster variation (in Euclidean space),

\begin{gather*}
\sum_{k=1}^K \sum_{\mathcal{C}(i) = k} || X_i - c_k ||^2,\\
\text{with } c_k = \overline{X}_k = \frac{1}{n_k} \sum_{\mathcal{C}(i) = k} X_i,
\end{gather*}
is minimized \citep{macqueen1967}. The resulting clustering $\mathcal{C}$ from $K$-means assumes the co-variance structure of the clusters is spherical, which works well for our purposes, as all pass locations are represented by circles in the images. 

Rather than searching over all possible partitions, Lloyd's algorithm is the standard approach used for determining $K$-means partitions $\mathcal{C}$:

\begin{enumerate}
    \item choose $K$ random points as starting centers $c_i, \hdots, c_K$,
    \item minimize over $\mathcal{C}$: assign each point $X_i, \hdots, X_n$ to its closest center $c_k$, $C(i) = k$,
    \item minimize over $c_i, \hdots, c_K$: update the centers to be the average points $c_k = \overline{X}_k$ for each $k = 1, \hdots, K$ clusters, 
    \item repeat steps (2) and (3) until within-cluster variation doesn't change.
\end{enumerate}

Rather than using random initialization for the cluster centers, we use the $K$-means++ algorithm to choose better starting values \citep{plusplus}. An initial point is randomly selected to be $c_1$, initializing the set of centers $C = \{c_1\}$. Then for each remaining center $j = 2, \hdots, K$: 

\begin{enumerate}
    \item for each $X_i$, compute $D(X_i) = \underset{c \in C}{\text{min}} ||X_i = c||$,
    \item choose a point $X_i$ with probability,
    $$
    p_i = \frac{D^2(X_i)}{\sum_{j=1}^n D^2(X_j) } 
    $$
    \item use this point as $c_k$, update $C = C \cup \{c_k\}$.
\end{enumerate}

Then we proceed to use Lloyd's algorithm from above with the set of starting centers $C$ chosen from a weighted probability distribution.  
According to this distribution, each point has a probability of being chosen proportional to its squared Euclidean distance from the nearest preceding center \citep{plusplus}.

The $K$-means++ algorithm is more appropriate to use for pass detection for two reasons. First, by definition, the initialized cluster centers are more likely to be spread further apart from each other in comparison to $K$-means. This means that the possibility of two cluster centroids falling within the same cluster/single pass location is greatly reduced.
Second, even though $K$-means++ is typically sensitive to outliers, our data does not present this issue, since all colors on the charts aside from the pass locations (e.g. the line of scrimmage and yardline markers) are already removed, as described above.  

To identify complete and intercepted passes, we simply perform $K$-means++ clustering on the appropriate segmented images with $K = n_c$ and $K = n_{int}$, respectively. 
While in theory we can do the same for the touchdown passes (i.e. perform $K$-means++ with $K = n_{td}$), the segmented image sometimes includes outliers, since the line of scrimmage and the touchdown pass trajectories have similar colors to the points representing catch locations for touchdown passes. 
Thus, we need to further process the corresponding segmented image to remove these unnecessary colored pixels. 
For this step, we use DBSCAN as we detail in the following section, and apply $K$-means++ to the resulting image.


One major difficulty in detecting the number of incompletions is that the number of incomplete passes shown on a pass chart image may not match the number given in the JSON data corresponding to an image, $n_{inc}$. 
This is most likely because out-of-bounds passes and spikes are not presented in the charts despite being counted as pass attempts. 
Our first step to solving this issue is performing $K$-means++ clustering on the segmented image for incompletions, with $K = n_{inc}$. Once we have all $n_{inc}$ cluster centers, we iterate through each cluster center and examine how far away the other cluster centers are. 
If two cluster centers are distinctly close to each other, one of following two cases is true:

\begin{enumerate}

\item Two cluster centers have been mistakenly detected for a single pass location.  This might happen if the metadata specifies that there are 21 incompletions, but the image only shows 20 (e.g. because one pass was out of bounds).  In this case, $K$-means++ will split a single pass location into two, in order to achieve the specified number of clusters $K$.

\item Two cluster centers have been correctly detected for two pass locations that are close to each other.  
\end{enumerate}

We can infer which of the two cases is true by comparing the within-cluster variation of each of these two clusters with the within-cluster variation of a {\em normal}, single pass location cluster. 
If the former is significantly smaller than the latter, then case (1) is detected and we reduce the number of incompletions shown in the image by one, otherwise, case (2) is detected and no additional action is required.
The result of this iterative process is a newly-adjusted number of incompletions, $n_{inc-adj}$. If $n_{inc-adj}=n_{inc}$ the process is terminated; otherwise we perform $K$-means++ clustering again with $K = n_{inc-adj}$.

\begin{figure}[!ht]
   \centering
    \includegraphics[width=.85\textwidth]{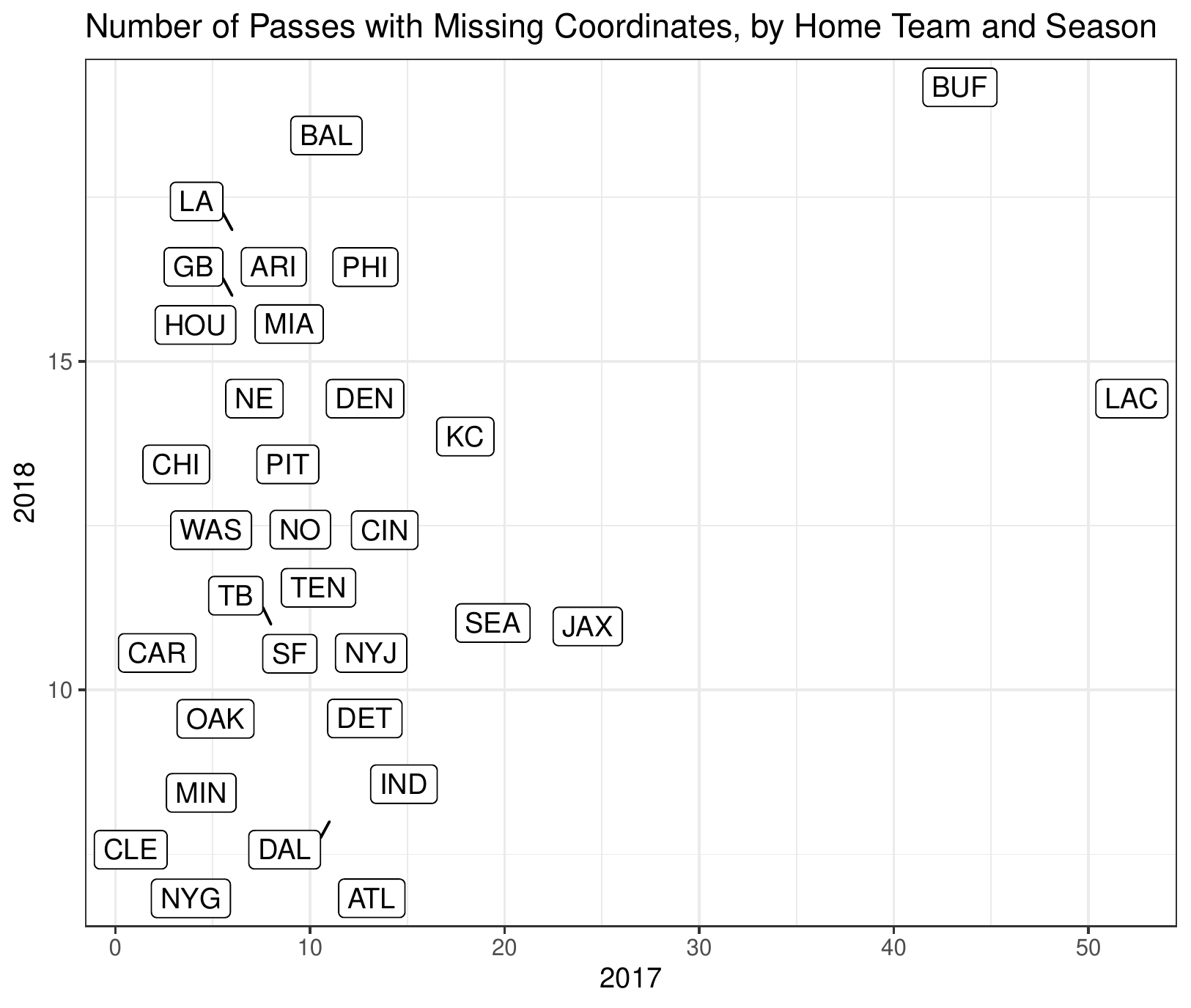}
    \caption{Number of missing pass attempt coordinates by home team and season.}
   \label{fig:missing}
\end{figure}

After obtaining the $(x,y)$ pixel locations of all cluster centers of a given pass type, we map these coordinates to real field locations relative to the line of scrimmage. For the number of incomplete passes whose locations could not be identified, $n_{inc} - n_{inc-adj}$, we populate these rows in the data with N/A values.  Only 2.8\% of pass locations in 2017 and 2.7\% of pass locations in 2018 had missing coordinates in the final output dataset.  These missing coordinates appear to happen more often for certain home teams (e.g. the LA Chargers and Buffalo Bills) than others (e.g. the Minnesota Vikings and Cleveland Browns).  With the exception of these outliers (Bills and Chargers in 2017), the number of missing pass locations by team is typically small, as shown in Figure \ref{fig:missing}.

\begin{figure}[!ht]
   \centering
    \includegraphics[width=.85\textwidth]{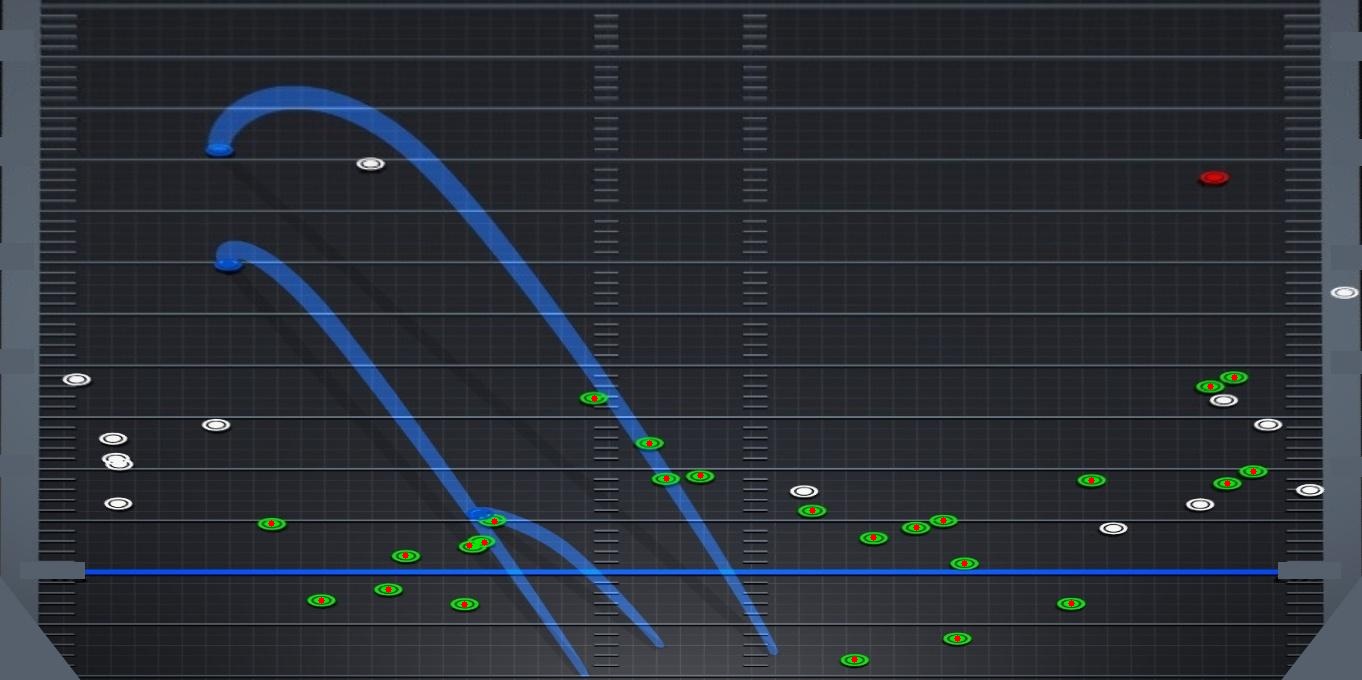}
    \caption{An example result of performing $K$-means on Nick Foles' complete passes from Super Bowl LII, with centers of each pass depicted in red.}
   \label{fig:kmeans}
\end{figure}

Figure \ref{fig:kmeans} shows the result of extracting pass locations using $K$-means++, and we mark the cluster centers in red. 
We note that $K$-means++ is able to detect overlapping pass locations as multiple passes.


\subsubsection{Removing Noise in Touchdown Images with DBSCAN}
\label{sec:dbscan}



Our segmented touchdown images require additional processing after color thresholding in order for $K$-means++ to correctly identify the pass locations. As mentioned above, this is because line of scrimmage is shown in blue, and because pass trajectories for touchdown passes are included in the image and shown in a similar color.  As a result, extraneous blue pixels (``noise'') are often present after image segmentation.

To address this issue, we use DBSCAN, a density-based clustering algorithm that identifies clusters of arbitrary shape for a given set of data points \citep{dbscan}.
DBSCAN is a non-parametric clustering algorithm that identifies clusters as {\em maximal} sets of density-connected points. Specifically, the DBSCAN algorithm works as follow:

\begin{enumerate}
    \item Let $X = \{x_1, x_2, ..., x_n\}$ be a set of observations (``points'') to cluster
    \item For each point $x_i$, compute the $\epsilon$-neighborhood $N(x_i)$; all observations within a distance $\epsilon$ are included in $N(x_i)$
    \item Two points, $x_i$ and $x_j$, are merged into a single cluster if $N(x_i)$ overlaps with $N(x_j)$
    \item Recompute the $\epsilon$-neighborhood $N(C_k))$ for each cluster $C_k$
    \item Two clusters, $C_k$ and $C_l$, are merged into a single cluster if $N(C_k)$ overlaps with $N(C_l)$
    \item Repeat steps 4-5 until no more clusters overlap
    \item If the number of points in a cluster is greater than or equal to a pre-defined threshold $\tau$ (i.e. if $|N(p)|\ge$ {\tt $\tau$}), this cluster is retained
    \item If the number of points in a cluster is less than a pre-defined threshold $\tau$ (i.e. if $|N(p)|<$ {\tt $\tau$}), this cluster is considered ``noise'' and discarded
\end{enumerate}

\begin{figure}[h]
\centering
\includegraphics[width=0.95\linewidth]{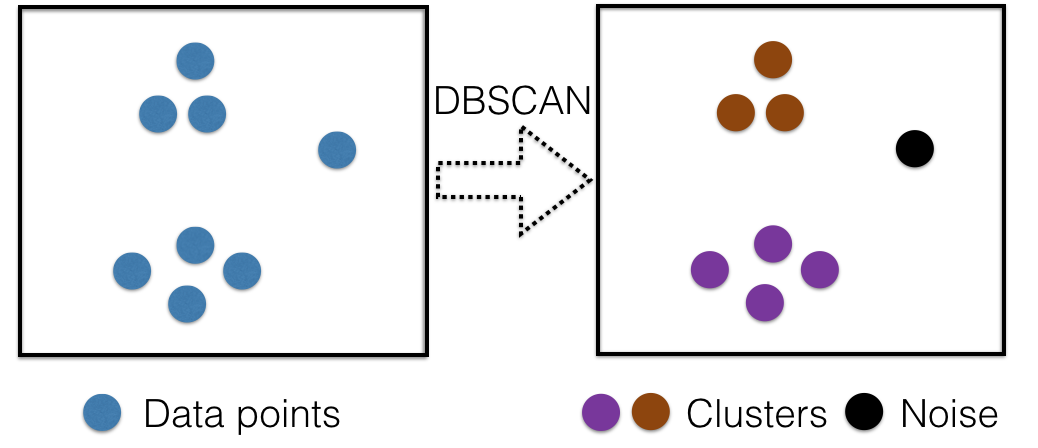}
\caption{Pictorial representation of DBSCAN's operations ($\tau$=3).}
\label{fig:dbscan}
\end{figure}

Even though DBSCAN does not require a direct specification of the number of clusters, it should be clear that the choice of $\epsilon$ and $\tau$ (often referred to as ``minPoints'') impacts the number of clusters identified.  Figure \ref{fig:dbscan} depicts a high level representation of DBSCAN's operations, with $\tau = 3$.

DBSCAN's ability to identify {\em noise} makes it particularly good choice for identifying the locations of touchdown passes. For our purposes, the observations are the individual pixels in the segmented image. 
To distinguish between actual pass locations and this noise, we use DBSCAN on these observations/pixels to find the $n_{td}$ highest density clusters, with $\epsilon = 10 $ and $\tau = n_{td}$. 
Then we remove from the image any pixels detected as noise or as not belonging to the $n_{td}$-top density clusters. 
We finally pass the resulting image to the $K$-means++ algorithm described above to obtain the raw locations. 



\begin{figure}[!ht]
   \centering
   \subfloat[][4 clusters identified by DBSCAN, circled in red. ]{\includegraphics[width=.47\textwidth]{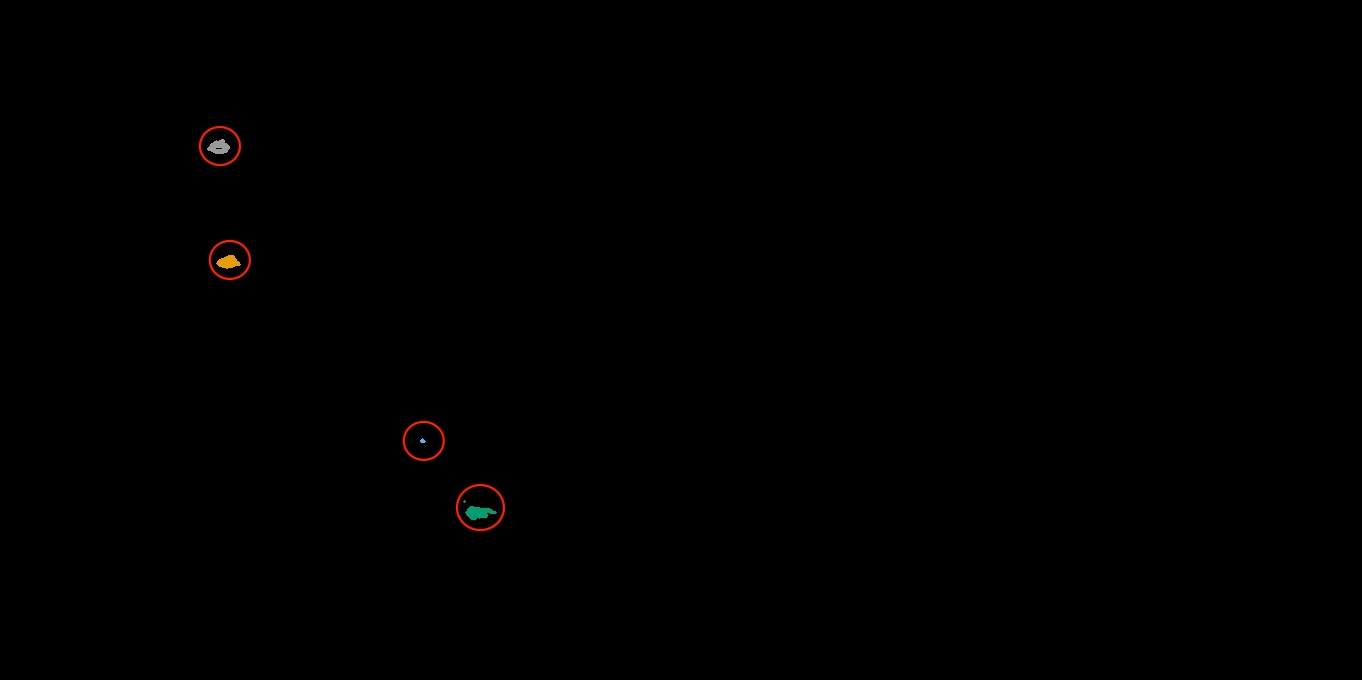}}\quad
   \subfloat[][$n_{td} = 3$ largest clusters]{\includegraphics[width=.47\textwidth]{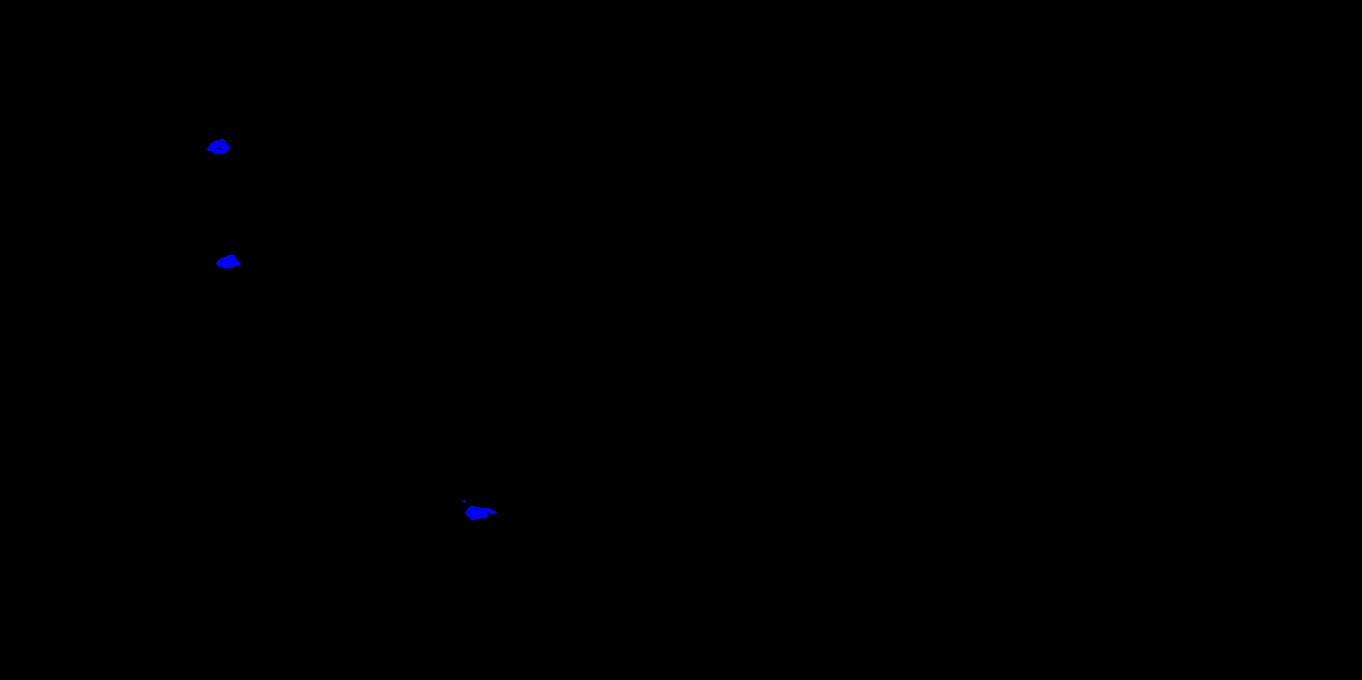}}\\
   \caption{The result of performing DBSCAN for touchdown pass detection. 
   }
   \label{fig:dbscan-output}
\end{figure}

Figure \ref{fig:dbscan-output} shows the output after applying DBSCAN on Figure \ref{fig:segmented}(d) to extract touchdown pass locations. 
In the original pass chart there are three touchdown passes, while as we see at Figure \ref{fig:dbscan-output}(a) DBSCAN has detected 4 distinct clusters (after removing the points identified by the algorithm as noise). 
Figure \ref{fig:dbscan-output}(b) shows the selection of the $n_{td} = 3$ most dense clusters that represent the touchdown pass locations.  This new version of the segmented image, with noise removed, is then used as input to the $K$-means++ approach described above.

\subsubsection{Output Data}

The resulting data from the pass charts cover the 2017 and 2018 regular seasons and postseasons. 
There are 27,946 rows containing data for 840 pass charts spanning 491 games, with 27,171 rows containing no missing values.
Of the cases with missingness, 33 are due to Next Gen Stats not providing pass charts for either team during a game. 
Appendix \ref{app:appendix-A} provides an overview of all variables provided by \method$\mbox{ }$and their descriptions, while Table 2 provides a breakdown of some basic statistics for the number of pass locations detected by season. 
Finally, Appendix \ref{app:appendix-B} contains a subset of the dataset for 10 of Nick Foles' passes in Super Bowl LII. 

\subsection{Validation}

In order to showcase the quality of the data collected by {\method}, we turned to the tracking data provided by the NFL for the Big Data Bowl competition mentioned earlier.  We provide two methods of validation of our algorithm:  First, an ``anecdote validation,'' where we visually inspect the pass locations obtained through {\method} and compare them to the pass locations in the NFL's tracking data (via the Big Data Bowl).  Second, we provide a method for linking {\method} passes to the corresponding play in the NFL's tracking data.

\subsubsection{Anecdote Validation}
\label{sec:anecdote}

We cannot compare all the data points collected from {\method} to the league-provided tracking data. 
For example, there are ambiguities in the tracking data when it comes to incomplete passes:  the NFL's tracking data does not specify the point at which the ball hits the ground and is rendered incomplete.  Therefore, we focus on completed passes (excluding touchdown passes), where the event of completion is clearly annotated in the tracking data. 

\begin{figure}
    \centering
    \includegraphics[width = \linewidth]{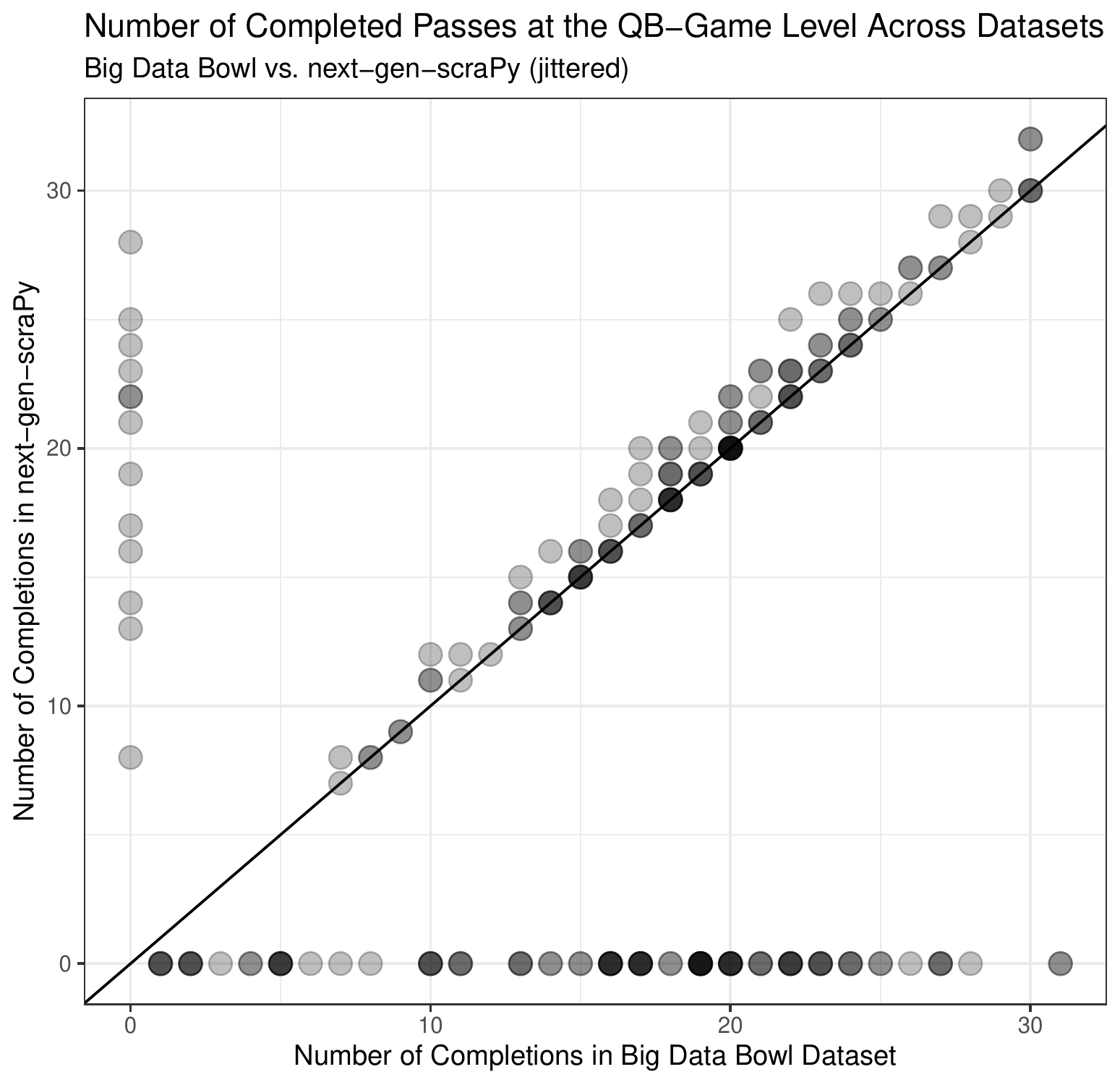}
    \caption{Number of completed passes (excluding touchdowns) at the QB-Game level in {\method} compared to NFL's tracking data from the Big Data Bowl.}
    \label{fig:number}
\end{figure}

Figure \ref{fig:number} shows the number of completed passes for each QB in each game across the {\method} and Big Data Bowl datasets, focusing on the first six weeks of the 2017 season, when the two datasets overlap.  There are small differences in the number of passes in each dataset for each QB in each game.  In games where both datasets contain completed pass locations for a given QB, {\method} always has at least as many completed pass locations as the Big Data Bowl Data, but the difference is never more than three passes; the two datasets agree (in terms of the number of completed passes for a QB) on most games for which they both have data.  Figure \ref{fig:number} also shows that while it's rare for a QB-game to be included in the {\method} dataset but not in the Big Data Bowl dataset, there are many QB-games that are included in the Big Data Bowl but whose passing charts are not available for inclusion in the {\method} dataset.

We further transform the tracking data coordinates in reference to the line of scrimmage in order to be directly comparable to the data obtained from {\method}.  We do this because the charts from which the {\method} data is sourced show coordinates \emph{relative to where the ball was snapped}, i.e. the line of scrimmage.  To reconcile the two coordinate systems, we first calculate the yards gained between the snap location (i.e. line of scrimmage) and the location at which the ball is caught in the tracking data.  Next, since the {\method} coordinates are shown with respect to the center of the football field, we calculate the horizontal distance (i.e. along the line of scrimmage) between where the ball was caught and the center of the field.

For a visual anecdote validation of our approach, we demonstrate the accuracy of the data obtained from {\method} in Section \ref{sec:accuracy} by plotting the coordinates from both datasets on a single chart for a single QB in a single game, repeating this for multiple QB-games.

\subsubsection{Linkage Validation}
\label{sec:linkage}

Using the same procedure to adjust the coordinates as above, we design a greedy one-to-one record linkage algorithm to link completed passes from the NFL's tracking data to completed passes in {\method} (excluding touchdown passes).  This can be thought of as an algorithm for defining a injection across the two data sources, with the caveat that there is no linear transformation that maps the points in one dataset to the points in the other.  That is, the locations across the two datasets deviate randomly, as shown in Figures \ref{fig:distance-difference} and \ref{fig:angle-difference}.

Our record linkage algorithm works as follows.  For each set of completed passes for a single QB in a single game:

\begin{enumerate}
    \item Compute the Euclidean distance between all pairs of coordinates across the Big Data Bowl and {\method} datasets.
    \item Identify the two closest pass locations and link them.
    \item Remove the distances corresponding to all other pairs involving the linked records from (2).
    \item Repeat (2) and (3) until there are no more pairs of records across datasets to link.
\end{enumerate}

Because we link the closest pair, and then the next closest remaining pair, and so on until there are no more pairs to link, we likely make some linkage errors for clusters of passes that are close together, and the resulting distances between linked coordinates are right-skewed (i.e. we likely make some linkage errors at the right-tail).  Results of this record linkage procedure are discussed in Section \ref{sec:accuracy}.

%% file: 3_Analysis.tex
In this section, we model pass completion percentage conditional on pass location for the NFL, individual passers, and team defenses. 
We focus on the field range between 10 yards behind the line of scrimmage to 55 yards in front of the line of scrimmage, since this is where almost all pass locations are located in the two-year span for which we have data. For each group (QB, team defense, or NFL-wide), our model has two main components: a two-dimensional kernel density estimate (KDE) for pass target locations and a generalized additive model (GAM) for completion percentage by location.

\subsection{NFL-Wide Models}
\label{sec:nfl-models}

Below, we describe two models for NFL passing:  (1) a league-wide pass target location distribution, estimated using kernel density estimation; and (2) a model for league-wide completion percentage by pass target location, estimated with a generalized additive model.  

It is important to recognize that both models use only observational data, and thus should only be used to describe what has happened in the past.  The observed data is biased in obvious ways:  QBs tend to target open receivers, defenses tend to cover high-leverage areas of the field more closely, and target locations depend greatly on the game situation (e.g. yards to first down).  None of this information is available in our dataset, but all of this information will influence on-field decisions that are made by QBs and teams.  

\subsubsection{Estimating the Distribution of NFL Pass Locations}

We estimate the league-wide pass location distribution via kernel density estimation (KDE).  KDE is a non-parametric approach for estimating a probability distribution given only observational data.  In the univariate case, a small probability density function (``kernel'') is placed over each observation $x_i$, and these kernels are aggregated across the entire dataset $x_1, ..., x_n$.  Let $K$ be the kernel function, $n$ the number of data points, and $h$ a smoothing parameter.  Then, the univariate, empirical density estimate via KDE is:  $\hat{f}_h(x) = \frac{1}{nh} \sum^n_{i=1}{K(\frac{x - x_i}{h})}$.

In our case, we use KDE to estimate the probability of a pass targeting a two-dimensional location $(x,y)$ on the field, relative to the line of scrimmage, $\hat{f}(x,y)$.  Density estimation via two-dimensional KDE follows a similar form: 

$$
\hat{f}_h(x,y) = \dfrac{1}{nh_xh_y} \sum^n_{i=1} \{K_x(\dfrac{x - x_i}{h_x}) K_y(\dfrac{y - y_i}{h_y})\}
$$

\noindent where $h_x$ and $h_y$ are bandwidths for $x$ and $y$, respectively, and $K_x$ and $K_y$ are the respective kernels.  Alternative definitions use joint kernels $K_{x,y}$ or identical kernels $K$, but for our purposes, the above definition suffices.

We use a bivariate normal kernel $K$ for our estimation (which can be decomposed into independent kernels $K_x$ and $K_y$), and Scott's rule of thumb heuristic for the bandwidth \citep{mass}.\footnote{$\hat{h} = 1.06 \times \text{min}(\hat{\sigma}, \frac{IQR}{1.34}) \times n^{-1/5}$, where $\sigma$ is the standard deviation and $IQR$ is the interquartile range of the values in the corresponding dimension}  We bound the KDE within the rectangular box described above:  10 yards behind the line of scrimmage, 55 yards in front of the line of scrimmage, and between both sidelines.  The resulting KDE gives us an empirical estimate of the pass target locations for the entire NFL. 

An alternative approach would be to obtain a two-dimensional histogram using a grid over the field and estimating the number of passes that were thrown in each of the grid cells. 
However, this approach has some limitations, including the histograms' sensitivity to the anchor point \citep{silverman1986density}. 
Furthermore, the estimates obtained from overlaying a grid over the field are essentially a discrete approximation of a continuous surface. 
Two points ($x_1, y_1$) and ($x_2, y_2$) that belong to the same grid, will have the same probability of being targeted with a pass since they belong to the same grid cell. 
Of course, this problem can be minimized by having the grid cells as small as possible (e.g., 0.5 yards each cell side), but then, sample size concerns limit the interpretability of the resulting density estimate, since we will end up with several empty cells and a noisy estimation. 
KDE provides a continuous approximation over the surface, smoothing the differences between neighboring points.

\subsubsection{Estimating the NFL Completion Percentages Surface}

We use generalized additive models (GAMs) to model the probability $P(\mbox{Completion} | x,y)$ of a pass being completed given the targeted location $(x,y)$ of the field \citep{hastie1990monographs}. 

GAMs are similar to generalized linear models (GLMs), except where the response variable is linearly associated with smooth functions of the independent variables (via some link function, e.g. the logit for a binomial response).  This is in contrast to typical GLMs (e.g. logistic regression), where the response variable is linearly associated with the independent variables themselves (via some link function). 
Formally, if $y$ is the dependent variable, and $x_i,~i\in \{1,\dots,n\}$ are the independent variables: 

$$
y = g(\alpha_0 + f_1(x_1)+f_2(x_2)+\dots+f_n(x_n))
$$
where $g^{-1}$ is the link function for the model. 
For binary data, as in our case, the link function is the logit function, $g^{-1}(z) = \log (\dfrac{z}{1-z})$.
The functions $f_i$ can take many forms, but can be thought of as smoothers between the dependent variable $y$ and the independent variable $x_i$. 

For our model, we use two independent variables -- the vertical coordinate $x$ and the horizontal coordinate $y$ -- and an interaction between them $x \cdot y$. We posit that all three terms are necessary:  
First, $x$ represents the pass location down the length of the field, and intuitively should have some marginal effect on completion percentage.  For example, longer passes are more difficult to complete.  
Second, $y$ represents the pass location across the width of the field, and intuitively should have some (non-linear) marginal effect on completion probability.  For example, passes towards the sidelines are farther away, while passes to the middle of the field are closer, potentially affecting completion probability; target locations towards the middle of the field typically have more defenders in that area, potentially affecting completion probability.
Third, we include the interaction term, since we hypothesize that there is some joint non-linear effect on completion probability that is not described by the marginal terms alone.  For example, the completion probabilities of passes behind the line of scrimmage likely are not aptly described by the marginal terms alone; there is likely some joint relationship between $(x,y)$ and the completion probability.
Thus, we have: 

$$\log(\dfrac{P(Complete)}{P(Incomplete)}) = c_0 + f_x(x) + f_y(y) + f_{xy}(x,y)$$
We choose $f_x$, $f_y$ and $f_{xy}$ to be tensor product interaction smoothers, which work well in situations where both marginal and interaction effects are present \citep{mgcv}. 
Specifically, this type of smoothing function accounts for the marginal bases from the main effects when estimating the smoothing function for the interaction effect, and it allows for the functional ANOVA decomposition in the model equation above, which is more interpretable than alternative choices.  Cross-validation supports our choice of smoother, and visual inspection indicates that this smoother yields interpretable results that make sense in the context of football (see Section \ref{sec:results} for examples).

Alternative models to the GAM are not ideal for this situation.  Generalized linear models, for example, will not allow for non-parametric or smooth relationships between the independent variables (distance from line of scrimmage, distance from center of field) and the response variable (completion percentage), and thus will yield a poor fit to the data.  Tree-based models (e.g. decision trees, random forests, gradient boosted trees, etc) partition the surface of the field into discrete areas, and will not yield the same smooth surfaces that we obtain from the GAM.  In short, the GAM is an ideal choice for the continuous nature of our response variable and the structure of our explanatory data; and it appropriately models and captures the relationship between our explanatory and response variables.

\subsection{2-D Naive Bayes for Individual Completion Percentage Surfaces}

For individual QBs or team defenses, we can directly apply the methodology used in Section \ref{sec:nfl-models} to subsets of the data corresponding to these groups.  As a result, we can obtain quick estimates of the pass target location for an individual quarterback and the completion percentage surface for an individual quarterback.  Similarly, for team defenses, we can take the subset of passes against this team (since opposing defense is included in the metadata described in Section \ref{sec:ngs-scrapy}), and examine the distribution of pass locations \emph{allowed} by a specific team, and the pass completion percentage surface \emph{allowed} by a specific team.

However, in doing so, we quickly run into sample size issues when estimating the completion percentage model.  While we had over 27,000 passes with which to estimate the league-wide models in Section \ref{sec:nfl-models}, a typical NFL team only attempts between 400 and 700 passes in an entire season, and the number of attempts for individual QBs can be even lower than that.  Moreover, in specific areas of the field that are less frequently targeted, some QBs may only attempt a handful of passes to that area of the field over the course of an entire season.  As such, any model built using these small samples of data for individual QBs and team defenses will likely overfit the data.

In response to this issue, we introduce a two-dimensional naive Bayes approach for estimating the completion percentage surfaces for individual QBs and team defenses.  Our approach regresses the estimates for individual QBs and team defenses towards the league-average completion percentage \emph{in each area of the field}, and adaptively accounts for the sample size of passes in each particular area of the field.  Our model is described as follows:

\begin{itemize}
    \item Let $g$ be the group or player of interest (e.g. an individual QB or a team defense).  
    \item Let $\hat{f}_{NFL}(x,y)$ be the pass target location distribution for the NFL.  Let $\hat{f}_{g}(x,y)$ be the pass target location distribution for group $g$.  
    \item Let $\hat{P}_{NFL}(\mbox{Complete} | x,y)$ be the estimated completion percentage surface for the NFL. 
    \item Let $\hat{P}_{g}(\mbox{Complete} | x,y)$ be the estimated completion percentage surface for $g$.  
    \item Let $N_{Median}$ be the median number of pass attempts in the class of group $g$ (for QBs, the median number of passes attempted for all QBs; for team defenses, the median number of pass attempts allowed for all teams).  (This is a parameter that can be adjusted to give more or less weight to the league-wide distribution.)
    \item Let $N_g$ be the number of total passes attempted in the NFL and for $g$ (or against $g$, in the case of team defenses).  
    \item Finally, let $\hat{P}_G^*(\mbox{Complete} | x,y)$ be our 2-D naive Bayes estimate of the completion percentage surface for $g$.  
\end{itemize}

Then, 

$$
\hspace{-1in}
\hat{P}_G^*(\mbox{Complete} | x,y) = \frac{N_g \times \hat{f}_{g}(x,y) \times \hat{P}_{g}(\mbox{Complete} | x,y) + N_{Median} \times \hat{f}_{{NFL}}(x,y) \times \hat{P}_{{NFL}}(\mbox{Complete} | x,y)} {N_g \times \hat{f}_{{g}}(x,y) + N_{Median} \times \hat{f}_{{NFL}}(x,y)}
$$

In words, our approach goes through each location on the completion percentage surface for $g$ and scales it towards the league-average completion percentage surface from Section \ref{sec:nfl-models}, using a scaling factor of $N_{Median}$.
The above approach is Bayesian in nature:  The second terms, on the right hand side of the numerator and denominator, can be thought of as the {\em prior} estimate for the completion percentage. Then, as we accumulate more evidence (i.e. as $N_{g}$ increases), the first term is given more weight. 

Similar alternatives may also be ideal, depending on the context.  For example, we could instead shift the weight from the prior to the data in group $g$ instead of using a static weight of $N_{Median}$.  This is easily implemented, and we encourage interested readers to use the open-source code that we provide for these models and try their own approaches.\footnote{https://github.com/ryurko/next-gen-scrapy}

%% file: 4_Results.tex
In this section, we use the data collected and the modeling framework described in the previous section to analyze the league-wide passing trends and the performances of individual quarterbacks and team defensese passing. 

\subsection{Accuracy of Data Provided by \method{}}
\label{sec:accuracy}

As described in Section \ref{sec:anecdote}, we cannot compare \emph{all} of the data provided by \method{} to league-provided Big Data Bowl tracking data for several reasons:  (1) the pass locations for incompletions are not marked in the tracking data; (2) the tracking data provides only six weeks of data from 2017 (while we provide data from all weeks from the 2017 and 2018 seasons); and (3) passes from {\method} do not contain play identifiers or timestamps, so we cannot join them with their counterpart pass from the Big Data Bowl.

\begin{figure}
    \centering
    \includegraphics[width = \linewidth]{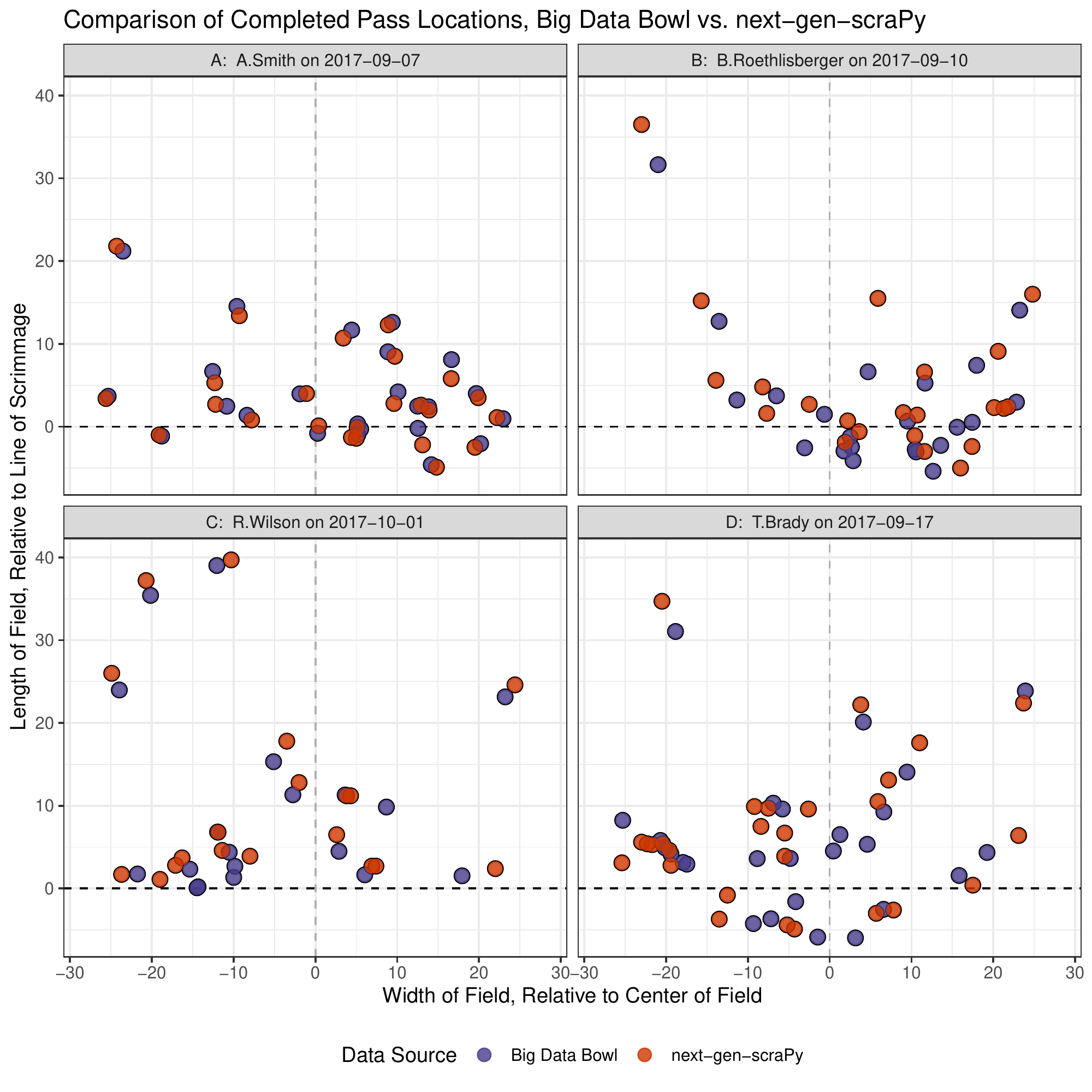}
    \caption{The pass locations obtained from {\method} closely match those from the NFL's player tracking data.}
    \label{fig:ngs_tracking}
\end{figure}

We can, however, provide examples of how our completed pass locations closely match those from the Big Data Bowl for individual QB-games.  Figure \ref{fig:ngs_tracking} visualizes the locations of the completed passes for four QB-games in the 2017 season:  (A) Alex Smith against New England on September 7th, (B) Ben Roethlisberger against Cleveland on September 10th, (C) Russell Wilson against Indianapolis on October 1st, and (D) Tom Brady against New Orleans on September 17th.  The points are very well-aligned, providing a validating example of the accuracy with which {\method} obtains raw pass locations by processing the images from Next Gen Stats.  

The small deviations for individual passes that we observe in Figure \ref{fig:ngs_tracking} could occur for several reasons.  For example, one source may mark the location of the ball, while another source may mark the location of the player catching the ball.  Since the Radio Frequency Identification (RFID) chips used to obtain the player tracking data are placed in players' shoulder pads, any catch that involves the receiver reaching out for the ball will have some small deviation between the ball and player locations.  Second, there may be differences in the timing of when these pass locations are tracked.  For example, the NFL tracking data annotates when a pass arrives in some places, and when it is caught in others.  It is possible that the data underlying the NGS passing charts uses one set of coordinates, while the NFL tracking data uses another.  If there is any time between these two events, and if the receiver moving, this may lead to a deviation in the coordinates across the two data sources.  Third, we use the center of field as the reference when calculating the horizontal distance that a pass travels in the Big Data Bowl Dataset, but it is possible that alternative approaches yield better results.  Fourth, there may be a small amount of error associated with our methods for identifying the locations of passes in the images.

\begin{figure}
    \centering
    \includegraphics[width = \linewidth]{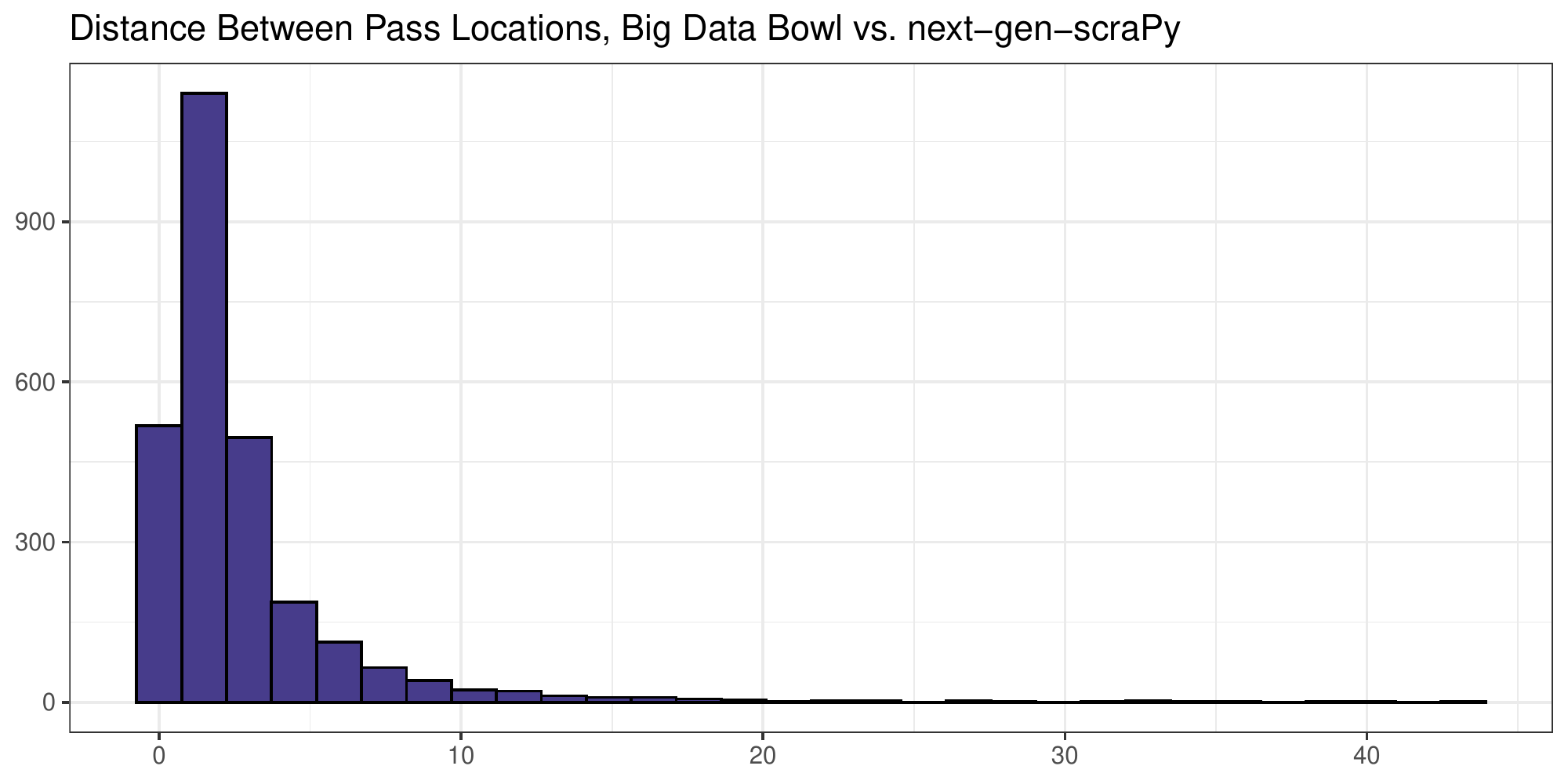}
    \caption{Difference in distances between completed pass locations in {\method} compared to NFL's tracking data from the Big Data Bowl.}
    \label{fig:distance-difference}
\end{figure}

\begin{figure}
    \centering
    \includegraphics[width = \linewidth]{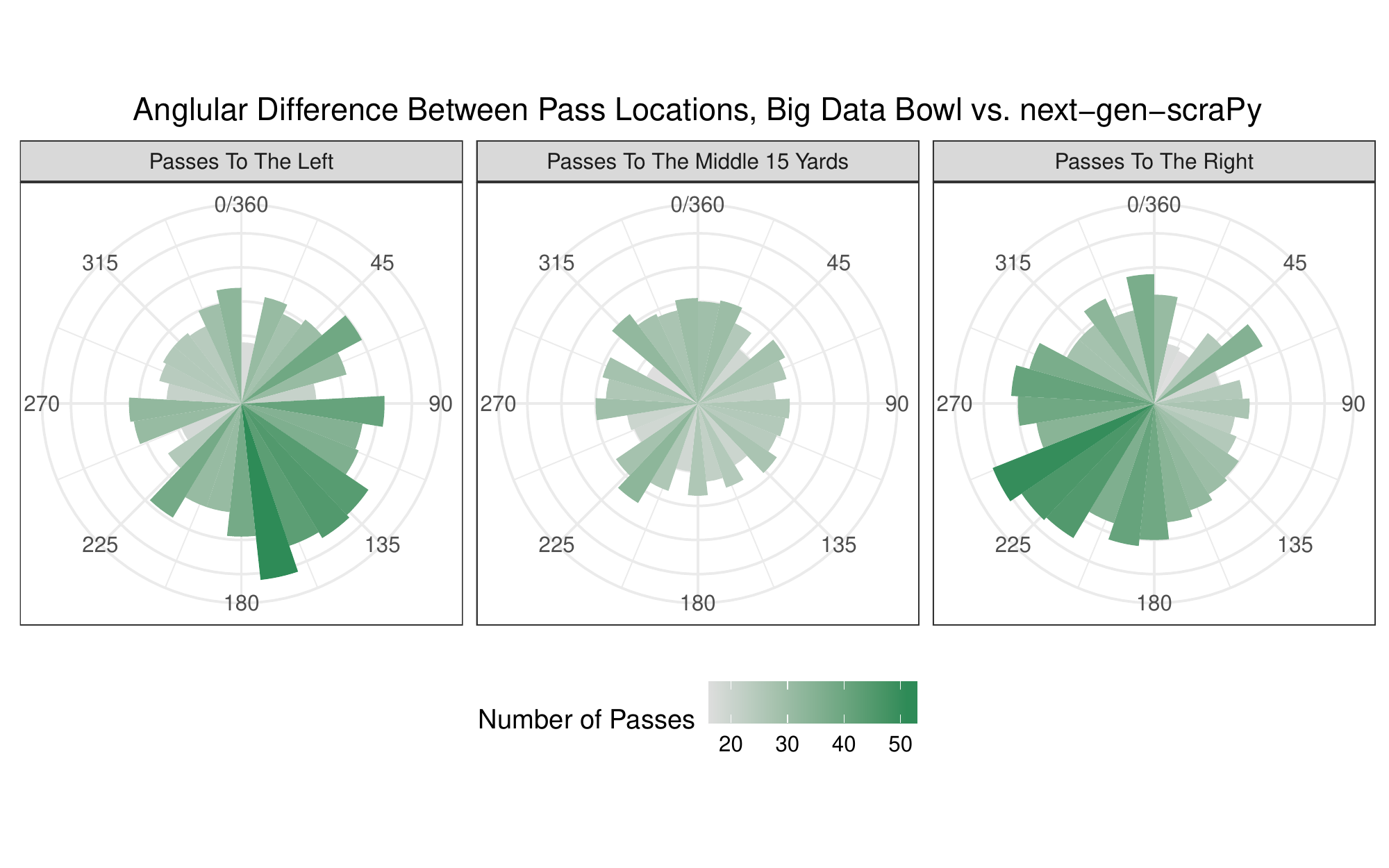}
    \caption{Angular difference between completed pass locations in {\method} compared to NFL's tracking data from the Big Data Bowl.}
    \label{fig:angle-difference}
\end{figure}

Circumstantial evidence for the first two reasons above can be found in Figures \ref{fig:distance-difference} and \ref{fig:angle-difference}.  Figure \ref{fig:distance-difference} shows that for most linked observations across the two datasets, the distance between the linked coordinates is small.  As discussed in Section \ref{sec:linkage}, the distribution of the distances is heavily skewed right (likely because we make linkage errors at the tails of our greedy record linkage algorithm).  The median deviation in the pass coordinates across the two datasets is only 1.7 yards.

Figure \ref{fig:angle-difference} provides some evidence that one data source marks the location of the ball, while another marks the location of the receiver.  As shown here, passes to the middle 15 yards of the field are not strongly associated with any angular difference across the two datasets.  However, for passes to the left side of the field, the Big Data Bowl Data pass locations tend to be closer to the line of scrimmage and more towards the right (i.e. towards the middle of the field) than those from {\method}.  Similarly, for passes to the right side of the field, the Big Data Bowl Data pass locations tend to be closer to the line of scrimmage and more towards the left (i.e. towards the middle of the field) than those from {\method}.  This might indicate that the Big Data Bowl coordinates correspond to the location of the player (in particular, the RFID chips in the players' shoulder pads), while the {\method} coordinates correspond to the location of the ball.

Finally, the pass locations from the NFL tracking data are only available for the first six weeks of the 2017 season.  It is possible that the NGS images have improved since then, and thus that the error associated with {\method} has decreased over time.  We look forward to re-examining this when additional, more recent tracking data is released by the NFL.


\subsection{League-Wide Passing Trends}

\begin{figure}[!ht]
    \captionsetup[subfigure]{labelformat=empty}
  \centering
  \includegraphics[width=.85\textwidth]{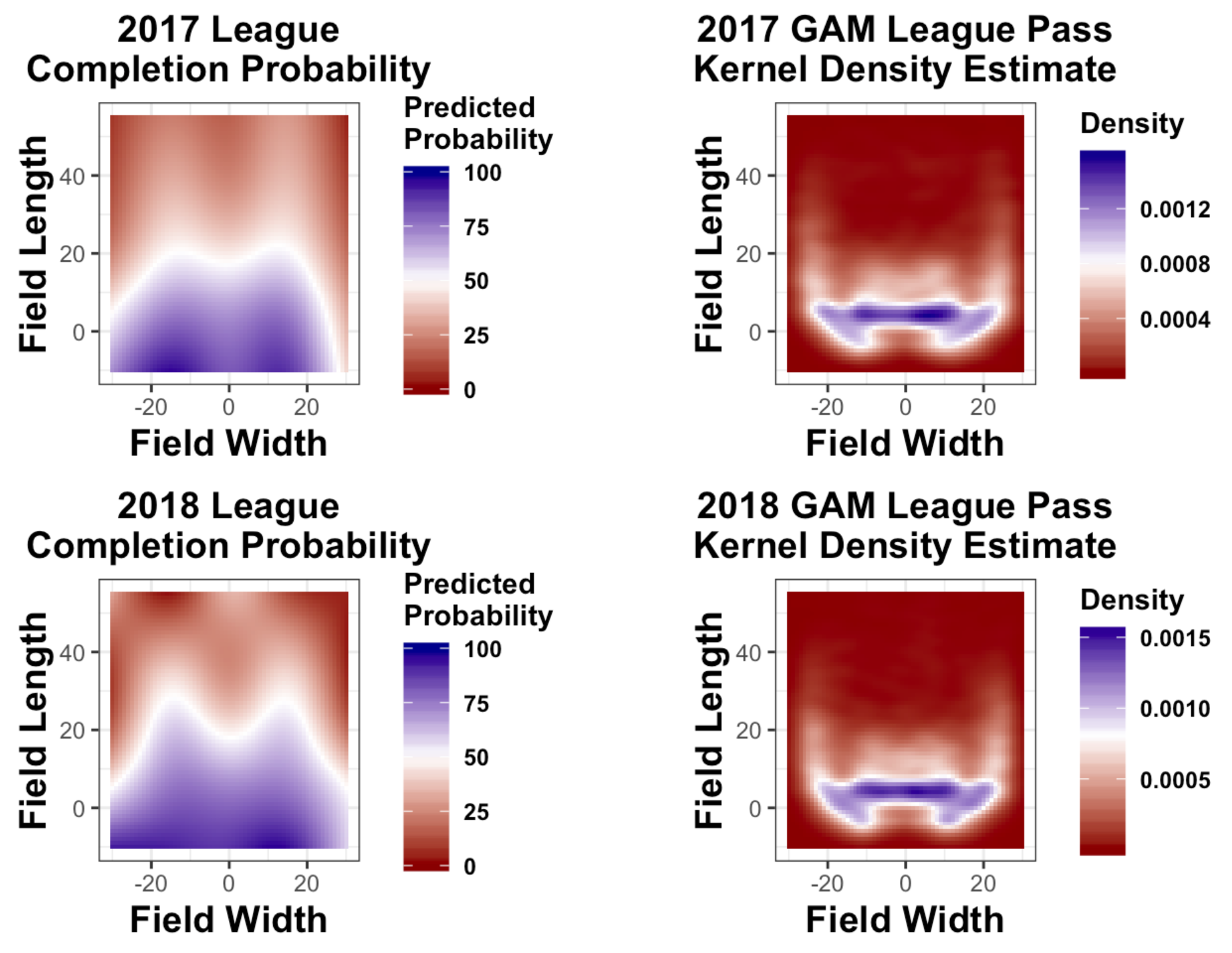}
  \caption{The individual components of our model: kernel density estimates for league-wide pass location probability (first column), and generalized additive models for league-wide completion probability (second column).} 
  \label{fig:league_wide}
\end{figure}

Figure \ref{fig:league_wide} shows the distribution of pass locations (relative to the line of scrimmage) and the results of the generalized additive model for predicting completion probabilities for the 2017 and 2018 NFL seasons. Comparing the two seasons, we can see that the distributions of pass locations are almost identical, with most passes are relatively short and near the line of scrimmage. 
Specifically, in 2017, 68.5\% of targets were within 10 yards down the field past the line of scrimmage, and 88.8\% were within 20 yards.  In 2018, 66.5\% of targets were within 10 yards, with 89.2\%  were within 20 yards. 
For targets further than 20 yards past the line of scrimmage, almost all were towards the sidelines. 
Only 2.96\% of targets past 20 yards in 2017 and 3.11\% in 2018 were in the middle of the field (i.e., $-13.33 \leq x \leq 13.33$). 

The predicted completion probability follows a similar trend. 
Passes closer to line of scrimmage have a higher completion percentage. 
The median predicted completion percentages for the 2017 and 2018 seasons are 73.3\% and 76.8\% respectively for passes within 10 yards of the line of scrimmage, while these numbers drop to 65.4\% and 69.7\% for passes within 20 yards. 
For passes more than 20 yards past the line of scrimmage, the median completion percentages fall drastically to 28.4\% and 28.0\% respectively. 
For these deeper passes, the median completion percentage was 30.4\% towards the middle and 26.5\% towards the sidelines in 2017, and 30.5\% towards the middle and 22.7\% towards the sidelines in 2018. 

This analysis is strictly retrospective, given the observational nature of our dataset, and does not include every game played throughout 2017 and 2018.  Potential biases in our dataset may result from the differences in play-calling for specific teams and QBs, differences in decision-making for specific teams and QBs, and the openness receivers and areas of the field to which passes are targeted.

\subsection{Team Defense Completion Percentage Allowed Surfaces}

\begin{figure}[!ht]
    \captionsetup[subfigure]{labelformat=empty}
   \centering
   \subfloat[][]{\includegraphics[width=.4\textwidth]{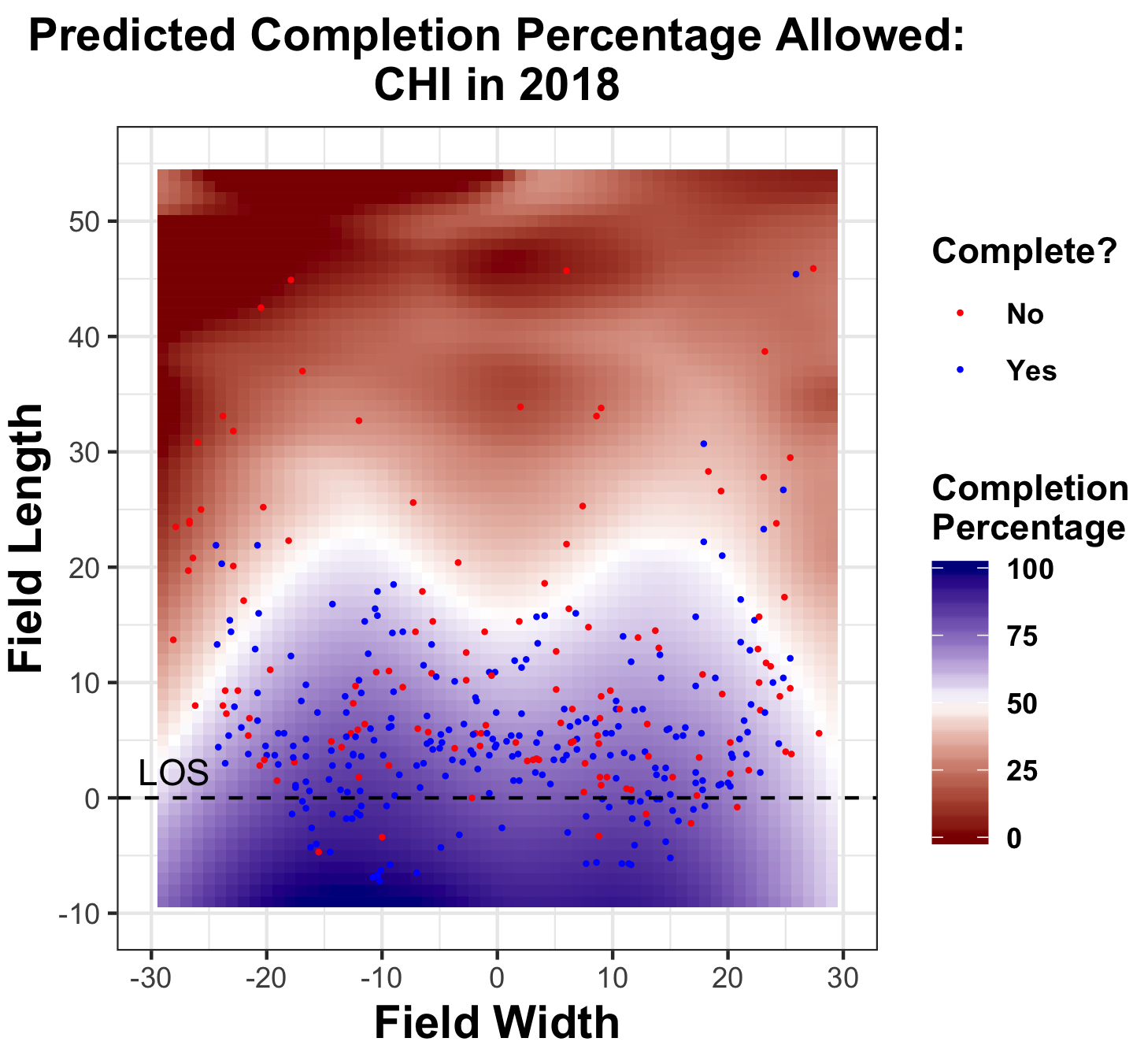}}\quad
   \subfloat[][]{\includegraphics[width=.4\textwidth]{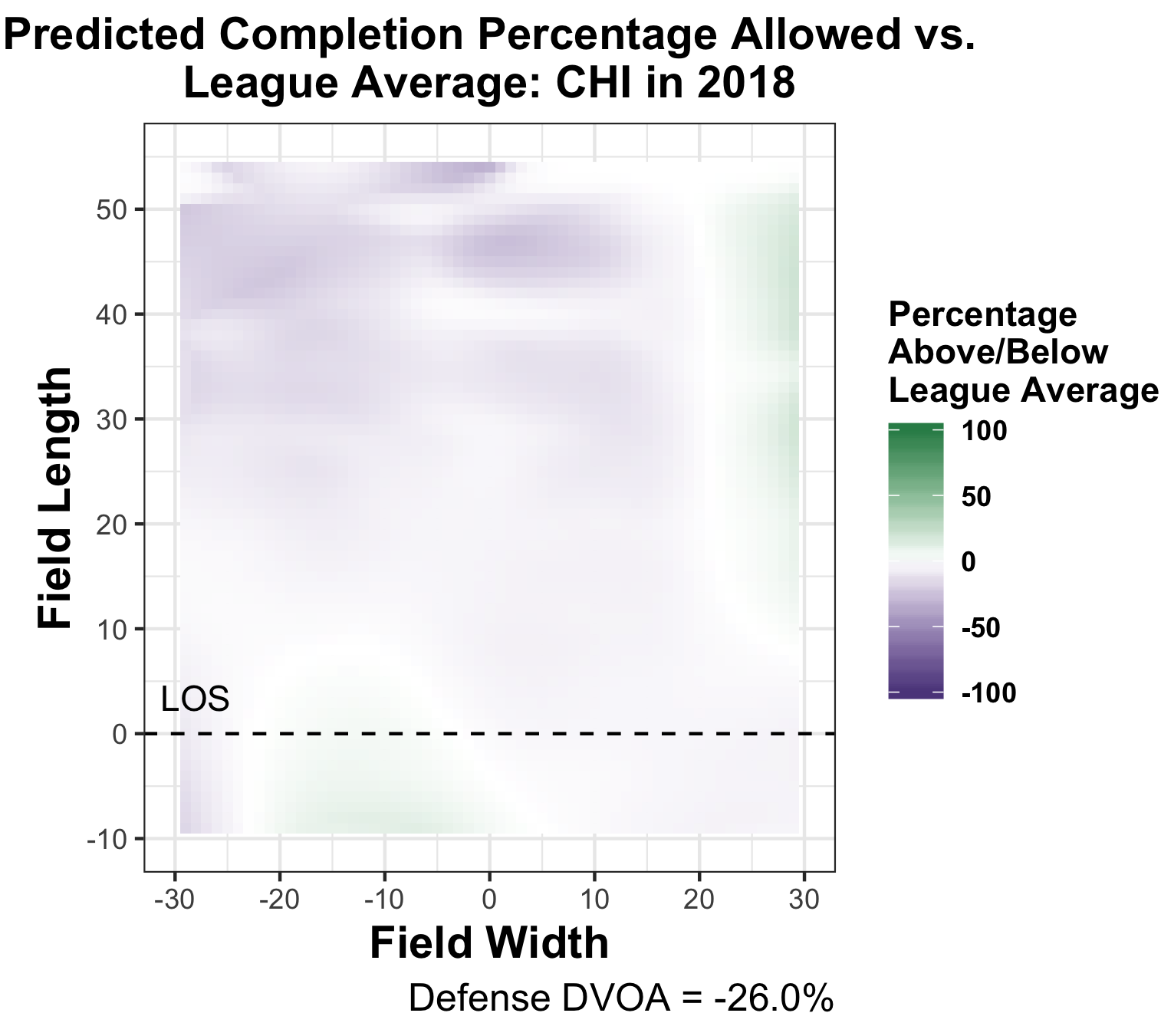}}\\
   \subfloat[][]{\includegraphics[width=.4\textwidth]{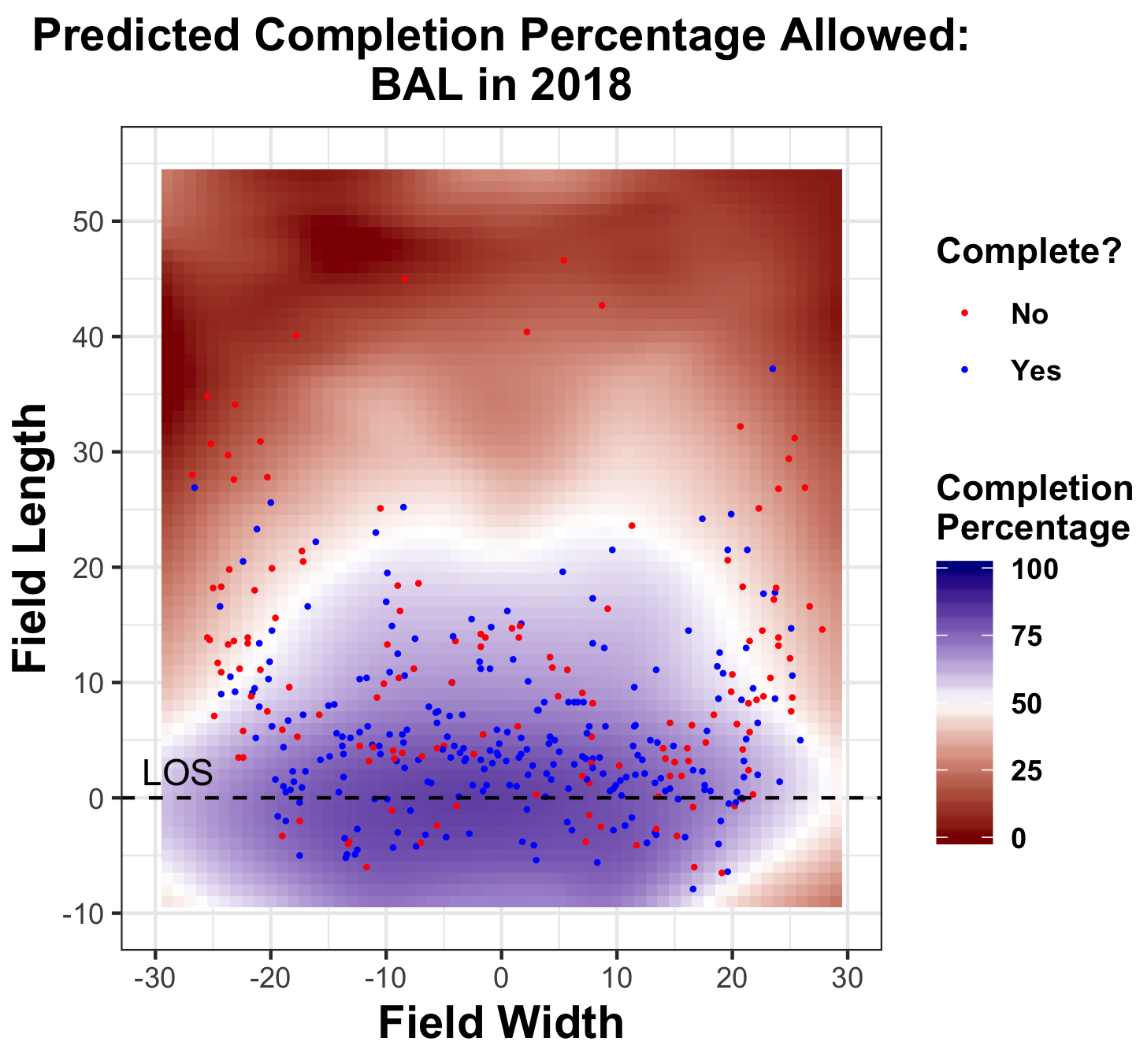}}\quad
   \subfloat[][]{\includegraphics[width=.4\textwidth]{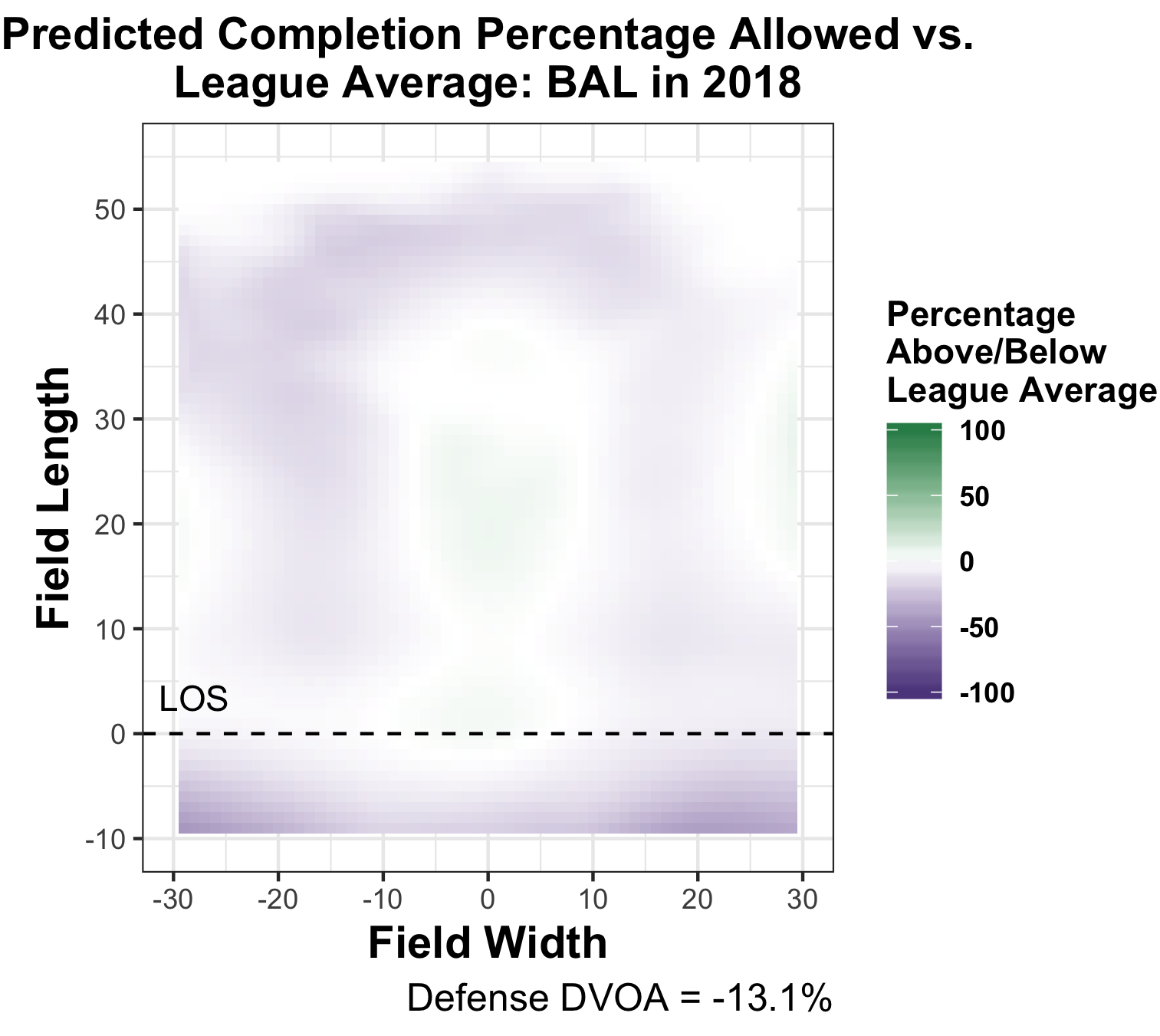}}\\
   \caption{Completion percentage allowed surfaces of the Chicago Bears and Baltimore Ravens defense, who have the highest Expected Points Contributed By All Defense according to Pro Football Reference.}
   \label{fig:good_team_d}
\end{figure}

\begin{figure}[!ht]
    \captionsetup[subfigure]{labelformat=empty}
   \centering
   \subfloat[][]{\includegraphics[width=.4\textwidth]{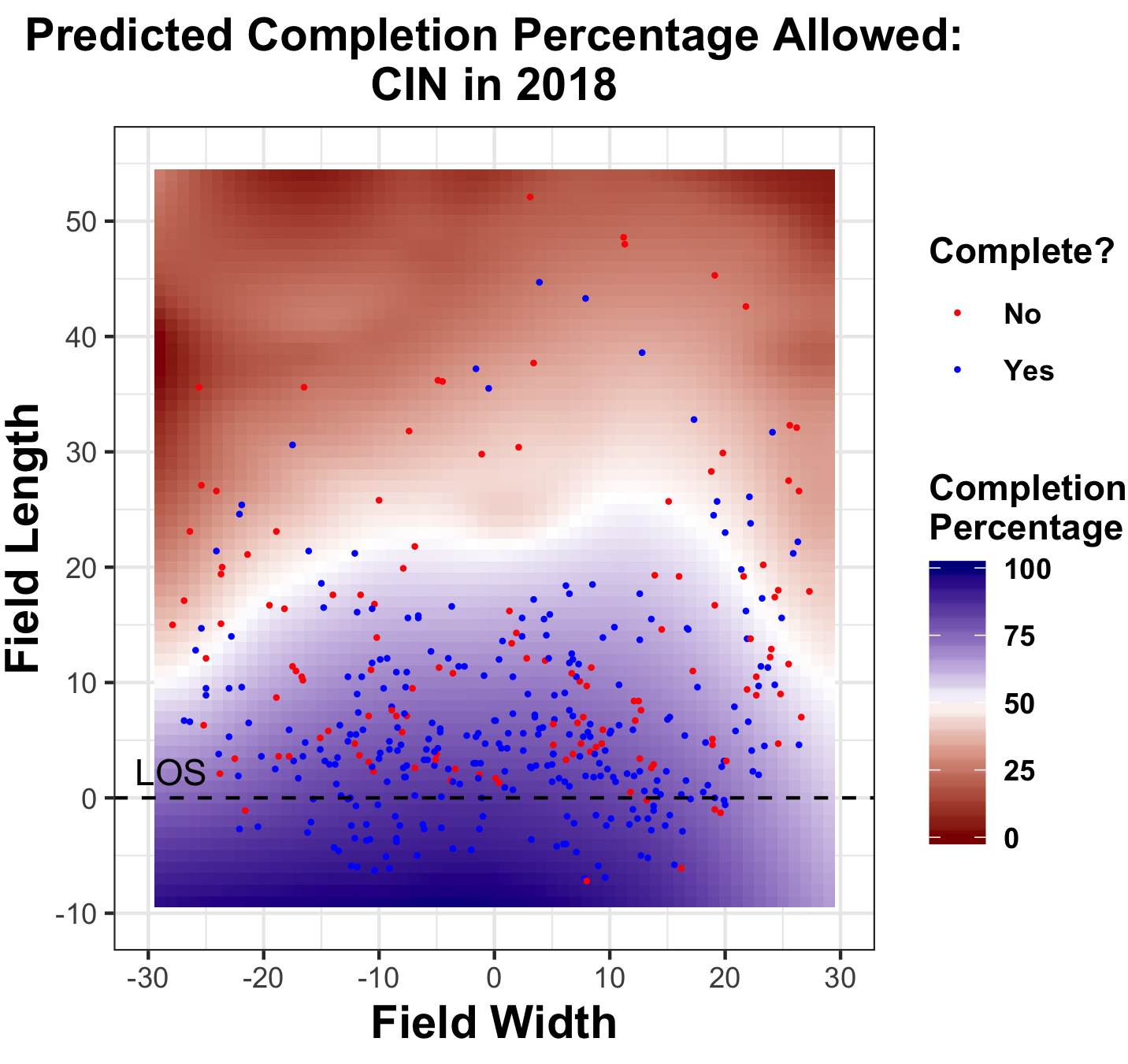}}\quad
   \subfloat[][]{\includegraphics[width=.4\textwidth]{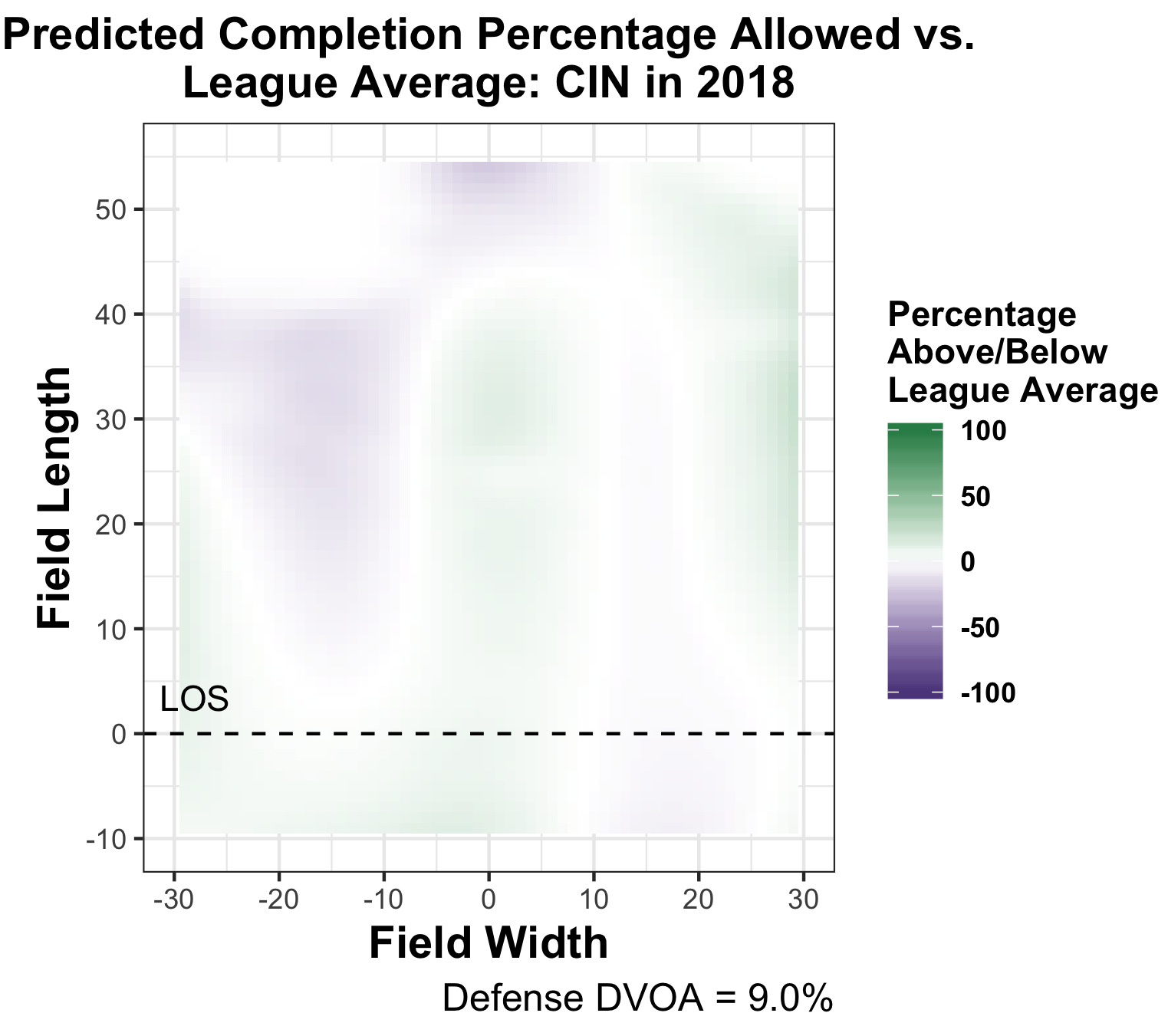}}\\
   \subfloat[][]{\includegraphics[width=.4\textwidth]{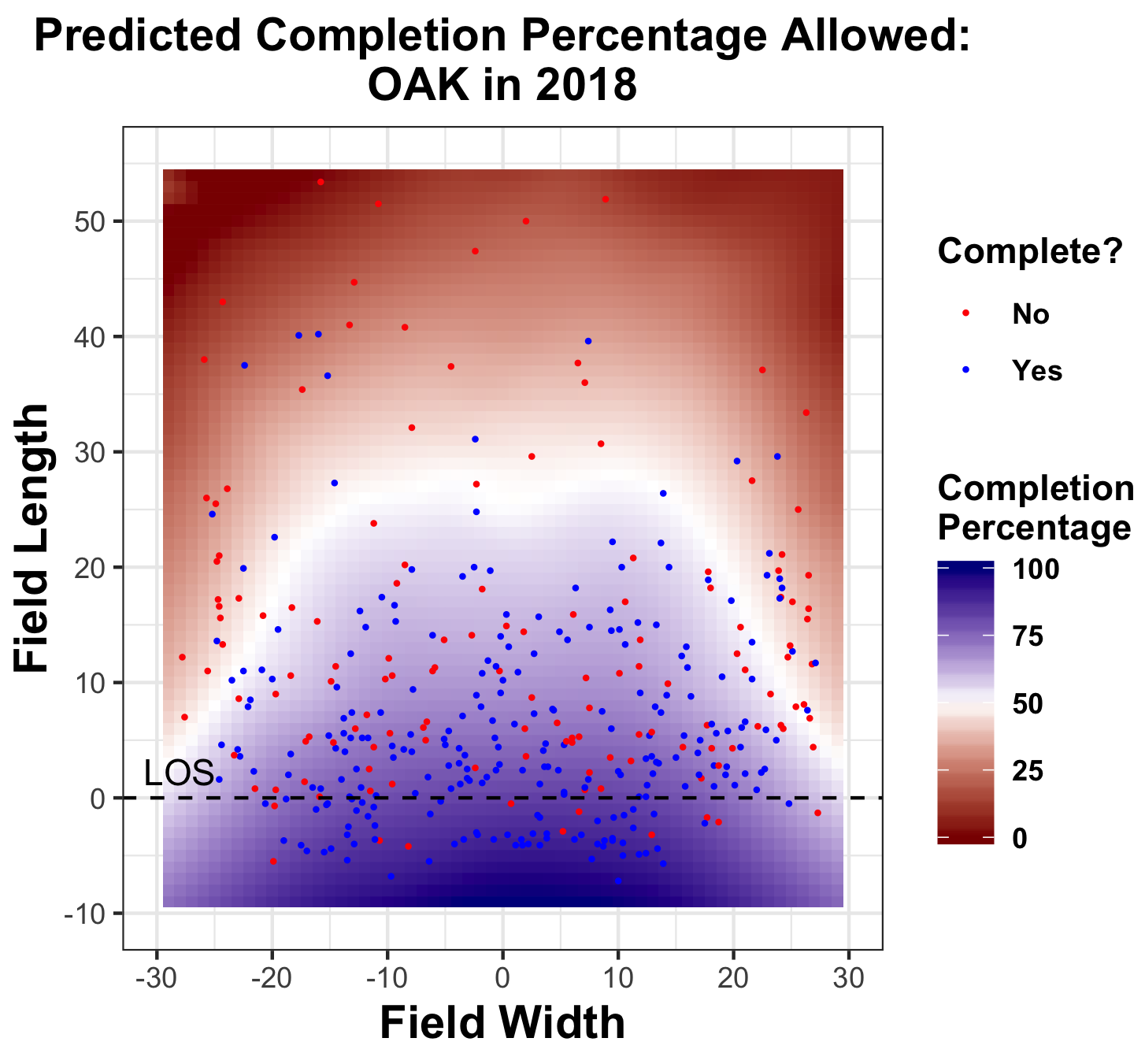}}\quad
   \subfloat[][]{\includegraphics[width=.4\textwidth]{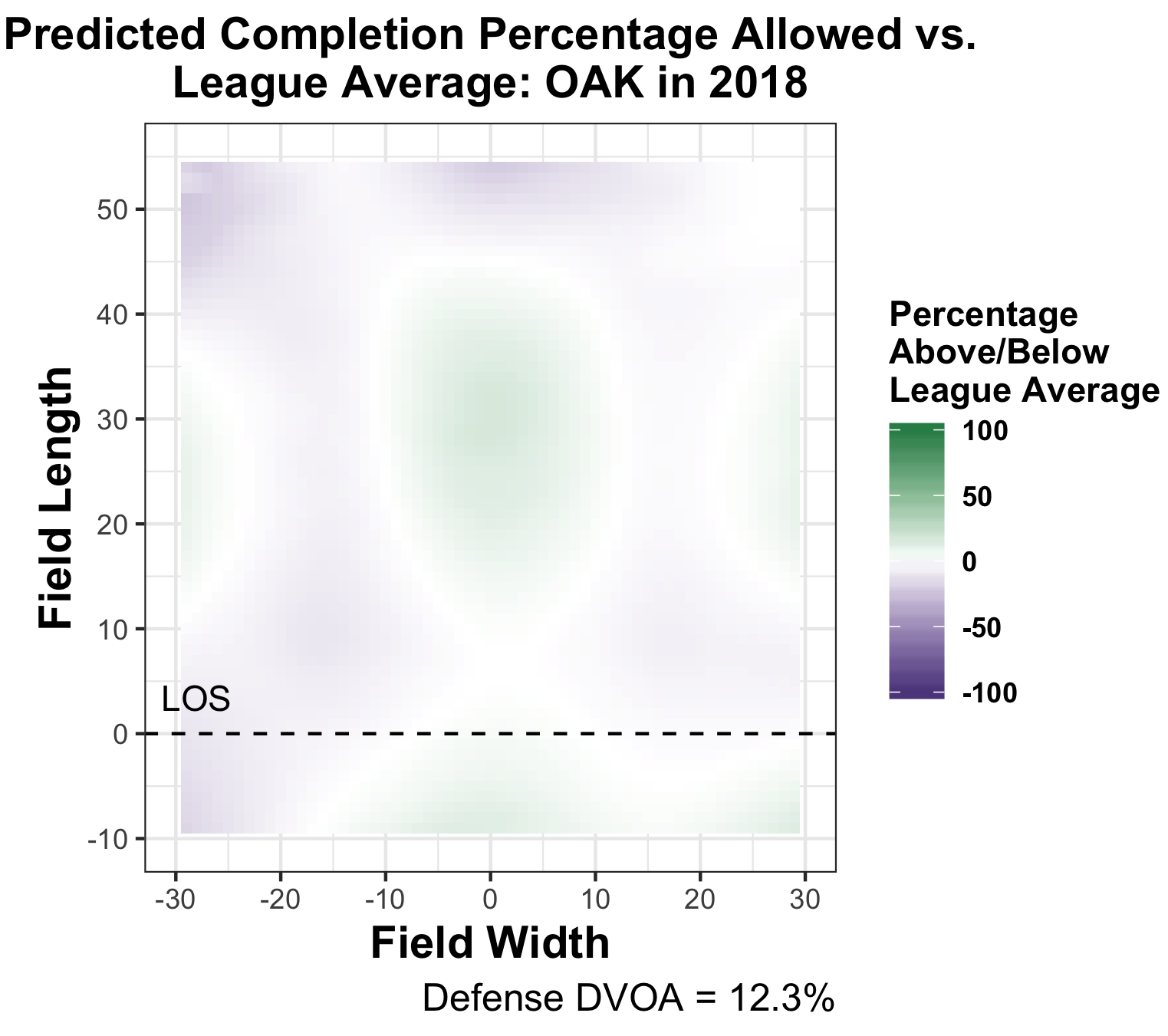}}\\
   \caption{Completion percentage allowed surfaces of the Cincinnati Bengals and Oakland Raiders defense, who have the lowest Expected Points Contributed By All Defense according to Pro Football Reference.}
   \label{fig:bad_team_d}
\end{figure}

Our framework also allows us to explore completion percentage allowed for each team defense by location of the field for an entire season. 
For this analysis, we compare our model output with Football Outsiders' Defensive Efficiency Ratings, specifically Defense DVOA (Defense-adjusted Value over Average). DVOA uses the idea of ``success points" to measure a team's success based on total yards and yards towards a first down for every play in an NFL season. With the Defense DVOA metric, $0\%$ represents league-average performance, a positive percentage signifies performance in a situation benefitting the opposing offense, and a negative percentage signifies performance benefitting the team's own defense \citep{FO_DVOA}.
In the 2018 regular season, the Chicago Bears and Baltimore Ravens had the first and third lowest Defensive DVOAs of $-26.9\%$ and $-14.2\%$, respectively. 
In Figure \ref{fig:good_team_d}, we present the predicted completion percentage from our model for the Chicago Bears and Baltimore Ravens. 
We present both the {\em raw} predictions (left column), as well as, relative to the league-average (right column). 
As we can see, in agreement with the Defensive DVOA metric, both teams are projected to allow lower completion percentage over the entire field. 
Furthermore, Figure \ref{fig:bad_team_d} depicts the same surfaces for the two defensive teams with the third and fifth highest Defensive DVOAs, namely, the Oakland Raiders at $12.3\%$ and Cincinnati Bengals at $9.0\%$. 
Compared to the results for the Bears and the Ravens, we can see that Bengals and Raiders are worse than average in completion percentage allowed over several parts of the field and particularly down the middle of the field. 

\subsection{Quarterback Completion Percentage Surfaces}

\begin{figure}[!ht]
    \captionsetup[subfigure]{labelformat=empty}
   \centering
   \subfloat[][]{\includegraphics[width=.4\textwidth]{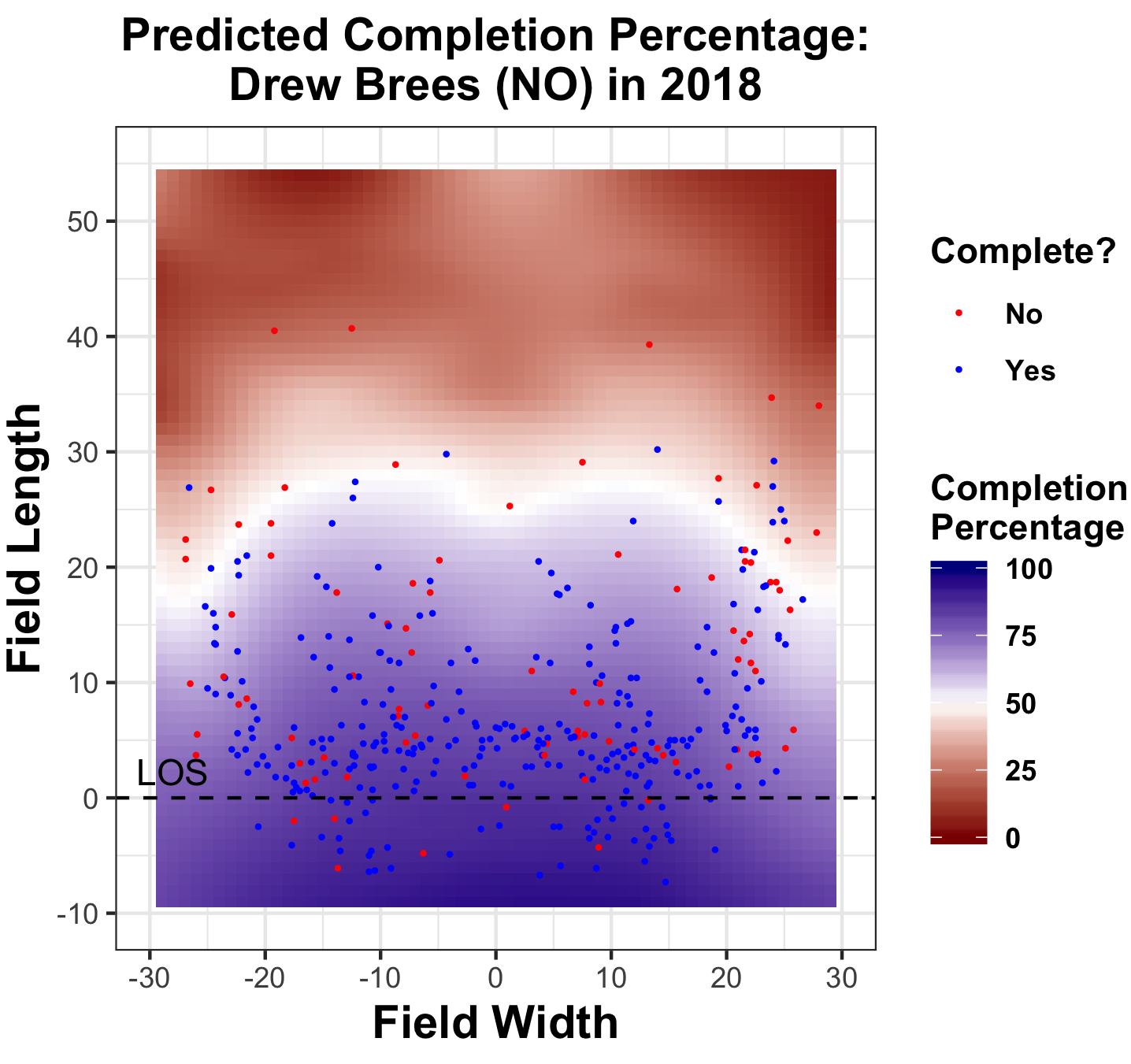}}
   \subfloat[][]{\includegraphics[width=.4\textwidth]{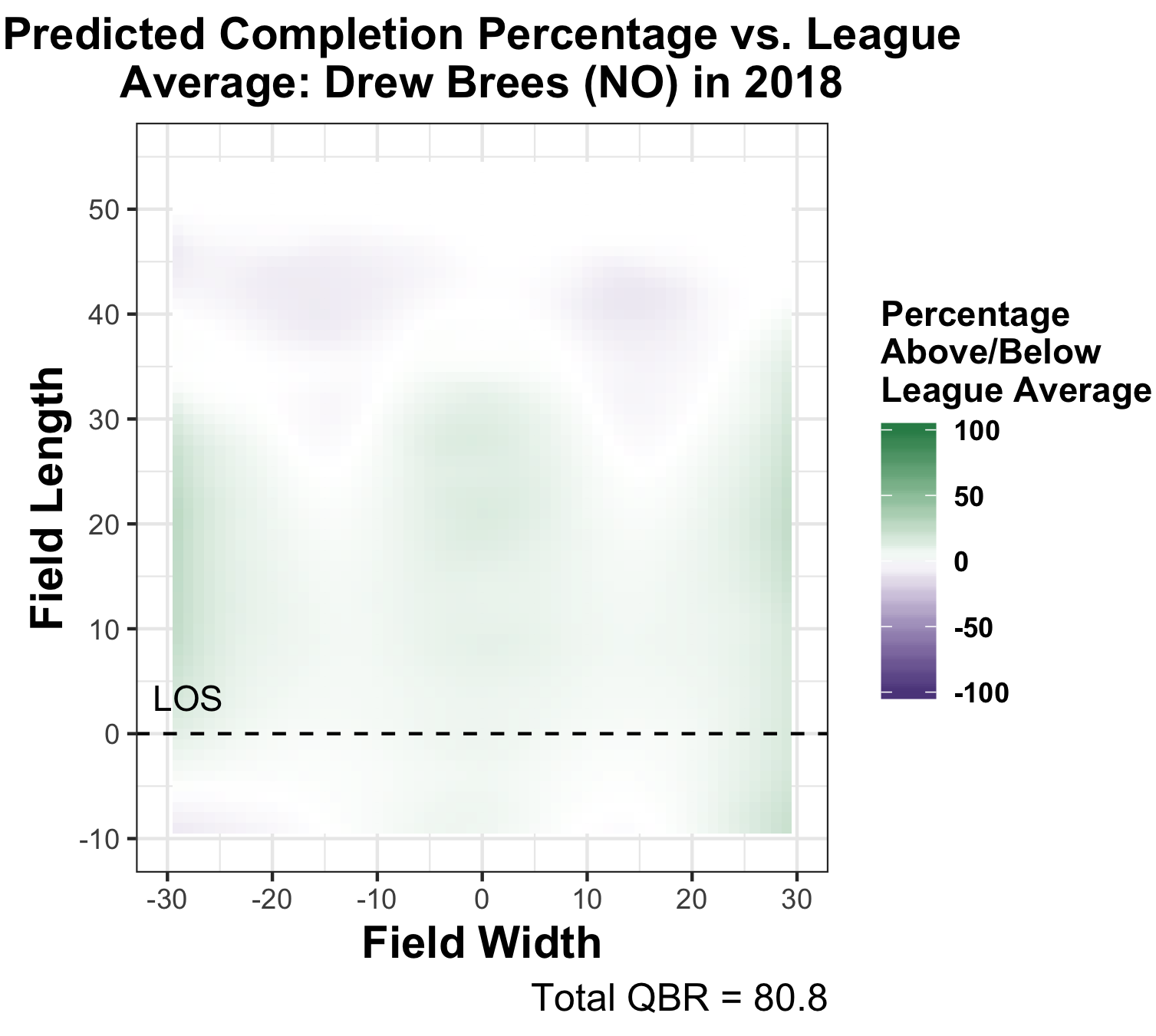}}\\
   \subfloat[][]{\includegraphics[width=.4\textwidth]{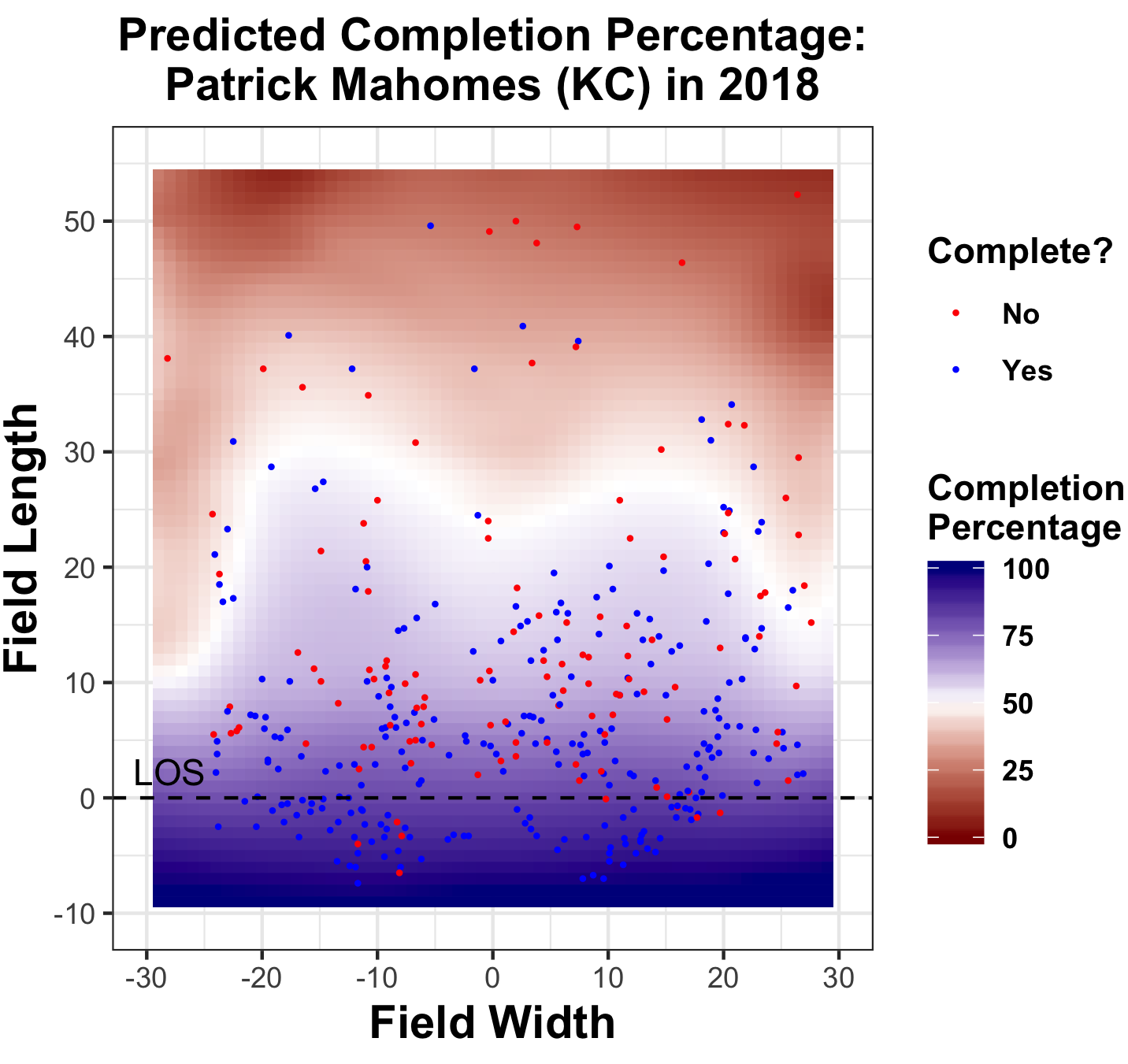}}
   \subfloat[][]{\includegraphics[width=.4\textwidth]{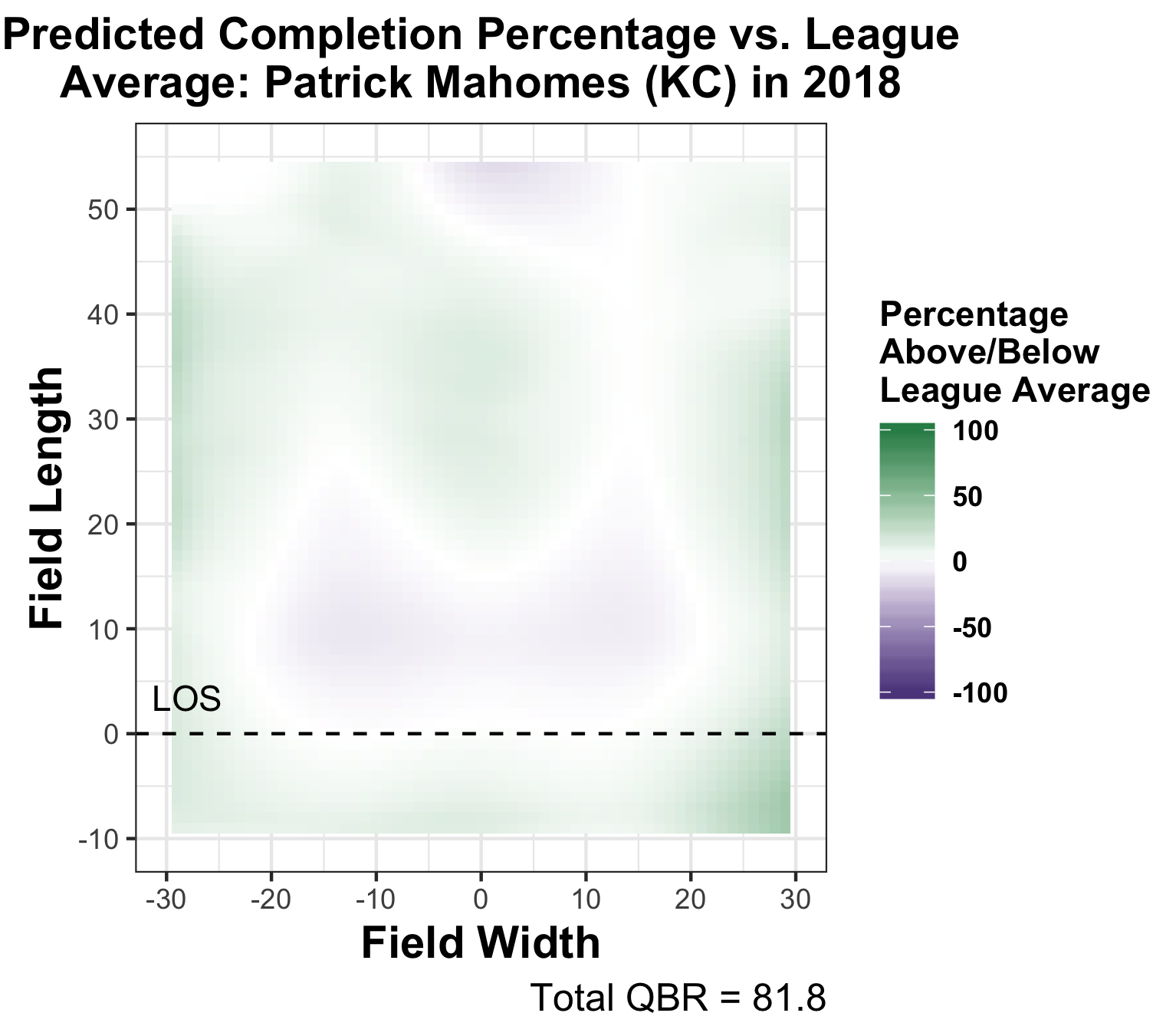}}\\
   \caption{Completion percentage surfaces for Drew Brees and Patrick Mahomes, the quarterbacks with the highest passer ratings in the 2018 regular season, according to Next Gen Stats.}
   \label{fig:good_qb}
\end{figure}

\begin{figure}[!ht]
    \captionsetup[subfigure]{labelformat=empty}
   \centering
   \subfloat[][]{\includegraphics[width=.4\textwidth]{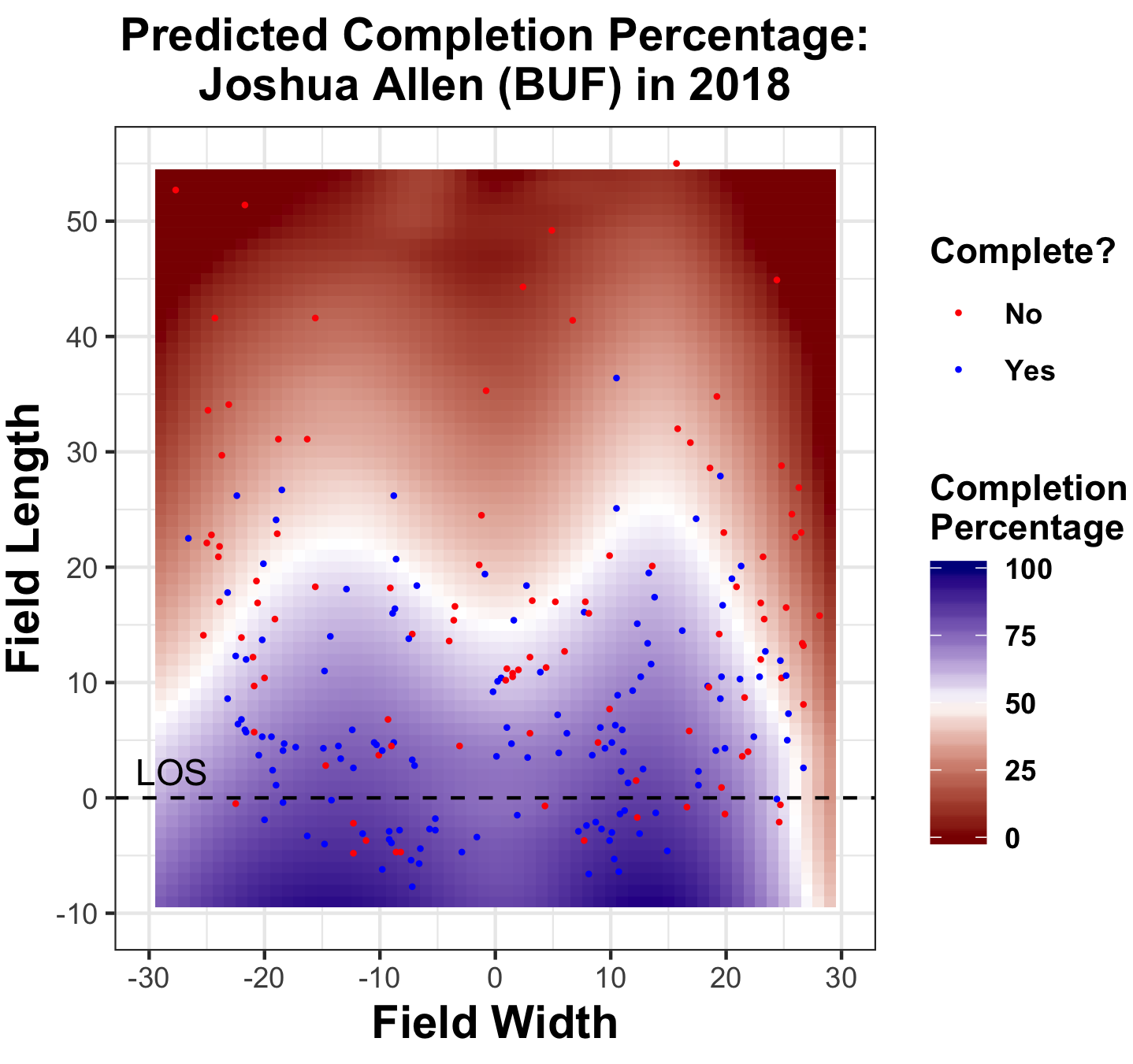}}
   \subfloat[][]{\includegraphics[width=.4\textwidth]{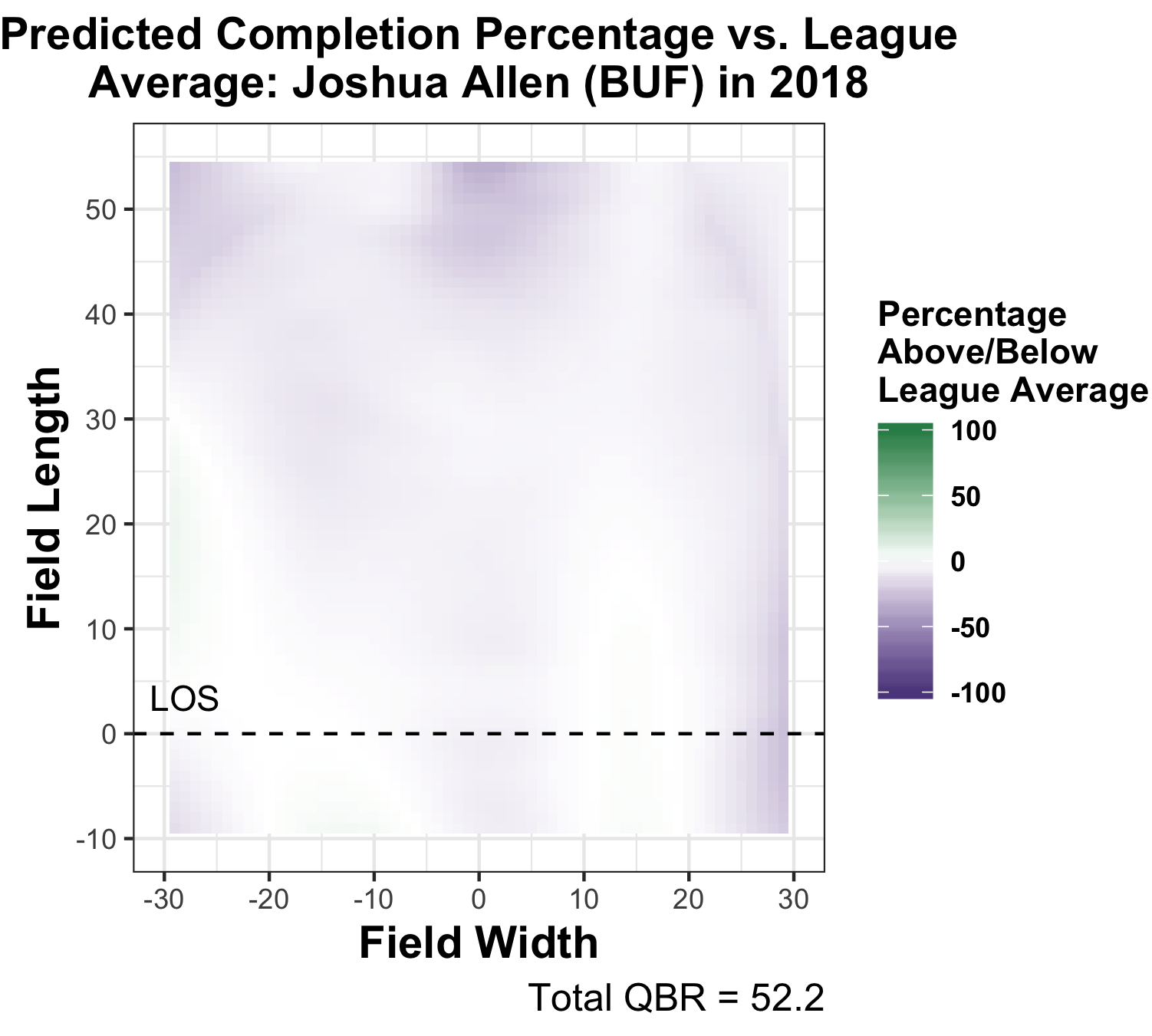}}\\
   \subfloat[][]{\includegraphics[width=.4\textwidth]{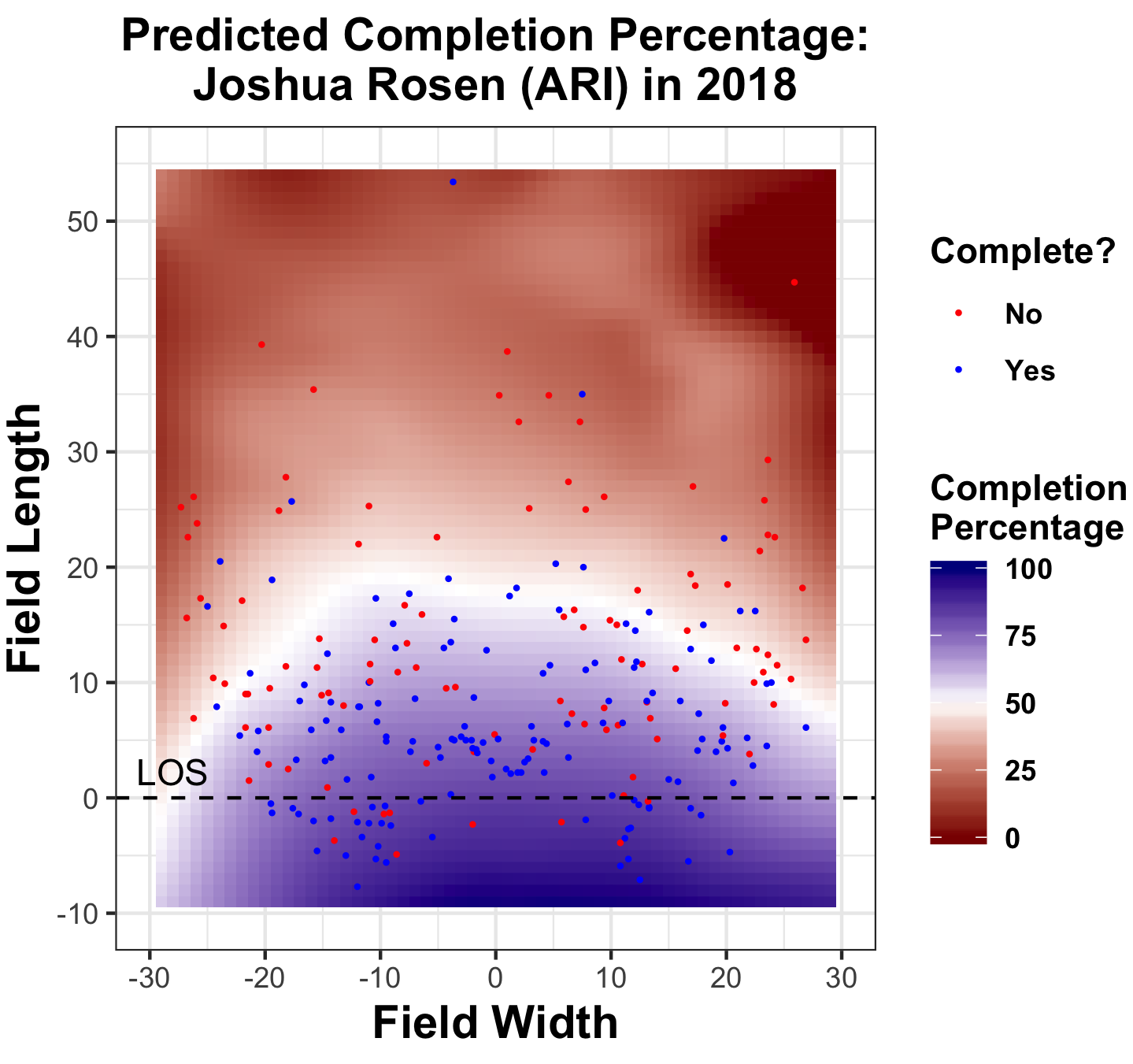}}
   \subfloat[][]{\includegraphics[width=.4\textwidth]{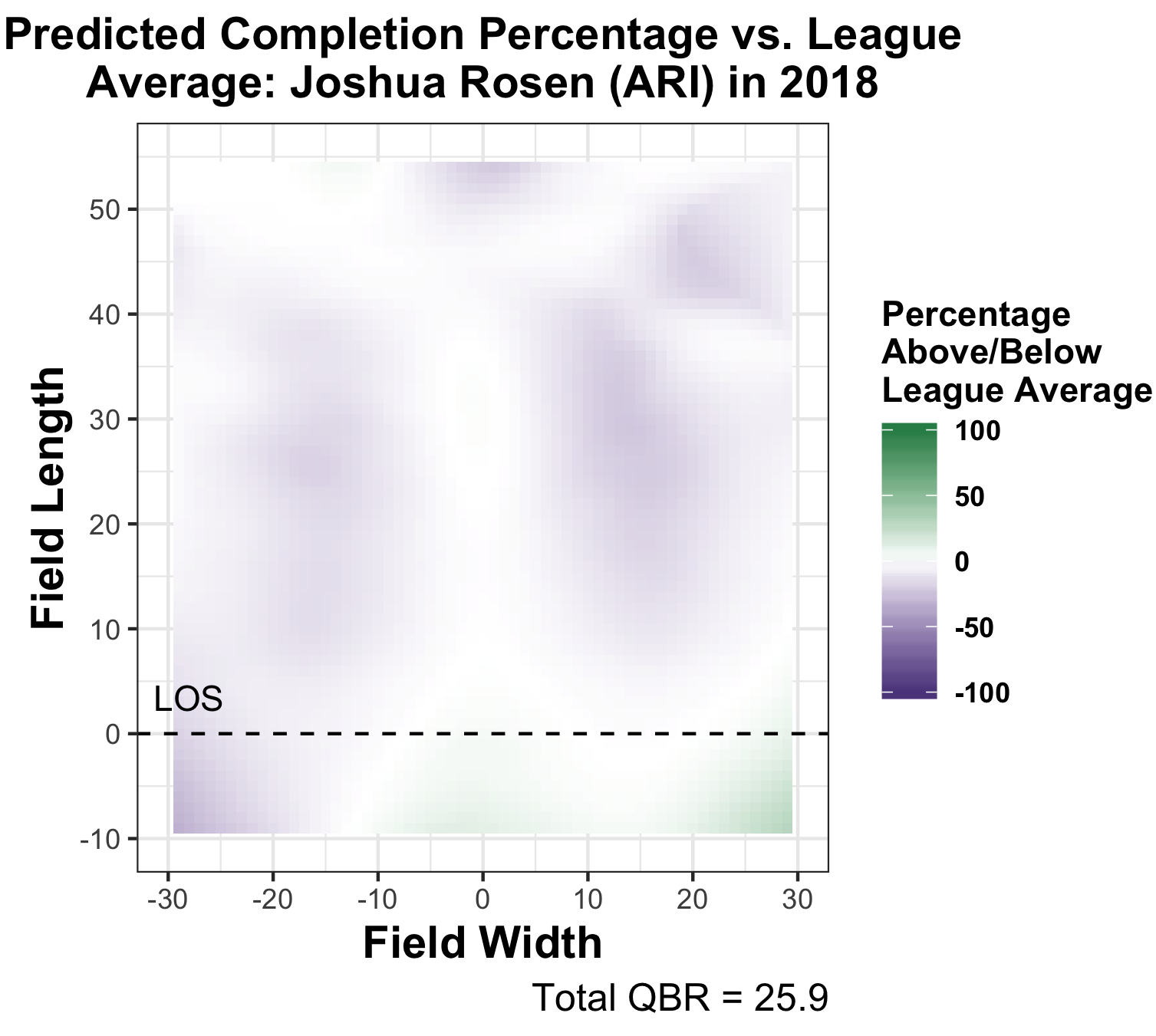}}\\
   \caption{Completion percentage surfaces for Joshua Allen and Joshua Rosen, the quarterbacks with the lowest passer ratings in the 2018 regular season, according to Next Gen Stats.}
   \label{fig:bad_qb}
\end{figure}

Our framework also allows us to examine patterns and make league comparisons at the quarterback level. We also compare our quarterback evaluations to ESPN's Total QBR metric, which describes a quarterback’s contributions to winning in terms of passing, rushing, turnovers, and penalties, as well as accounting for his offense's performance for every play \citep{ESPN_total_QBR}. 
In the 2018 regular season, the quarterbacks with the highest Total QBRs were Patrick Mahomes and Drew Brees, with total ratings of 81.8 and 80.8, respectively. 
Figures \ref{fig:good_qb} and \ref{fig:bad_qb} present the predicted from our model completion percentage by field location for the two QBs overlaid by the raw pass charts. 
When comparing Brees and Mahomes to league average, we find that both quarterbacks perform either equal to or above league average in almost possible target location on the field, indicated by the white (average) and green (above average) areas. On the other end of the spectrum, the quarterbacks with the lowest total ratings were Joshua Allen, ranked $24^{th}$ in the league, and Joshua Rosen, ranked last in the league at $33^{rd}$, with ratings of 52.2 and 25.9, respectively. 
Similarly, when comparing Allen and Rosen to league average, we observe that they performed below average (purple) across the field. 

\subsection{Completion Percentage Above Expectation}
\label{sec:cpae}

Using the estimated completion percentage surfaces for team defenses or individual QBs, we can obtain a single-valued summary for the performance of a QB or team defense in terms of completion percentage. 
In particular, we define the Completion Percentage Above Expectation of $g$ ($CPAE_g$) as the integral of the above-league-average completion surface for $g$, $\hat{P}_{g,league}^*(\mbox{Complete} | x,y)$, weighted by the spatial density of pass attempts for $g$, $\hat{f}_{g}(x,y)$:

\begin{equation}
    CPAE_g = \Int_{\mathcal{X}} \Int_{\mathcal{Y}} \hat{P}_{g,league}^*(\mbox{Complete} | x,y)\cdot \hat{f}_{g}(x,y)  \,dx\,dy
    \label{eq:cpae}
\end{equation}
where the double integral is over the two spatial dimensions of the field. 

We estimated the $CPAE_g$ for each QB with at least 100 passes for both seasons in our dataset, and Table \ref{tab:cpae} in Appendix \ref{app:appendix-C} presents the results.  From the table, we see that in the 2018 season, Drew Brees (+6.14\%), Ryan Fitzpatrick (+3.42\%), Nick Foles (+3.42\%), Russell Wilson (+3.39\%), Matt Ryan (+3.22\%), and Carson Wentz (+3.08\%) performed best by this measure; while Blake Bortles (-5.04\%), Jeff Driskel (-4.83\%), Josh Rosen (-4.54\%), and Casey Beathard (-4.37\%) performed worst by CPAE.

We similarly estimate $CPAE_g$ for each team defense in both seasons of the dataset, and Table \ref{tab:cpae_def} in Appendex \ref{app:appendix-D} presents the results.  Noting that negative CPAEs are better when evaluating team defenses (i.e. they reduce opponents' completion percentage), we see that in the 2018 season, the best pass defenses were Baltimore (-5.24\%), Chicago (-2.43\%), Los Angeles Rams (-2.35\%), Oakland (-2.12\%), and Kansas City (-1.98\%); while the worst pass defenses were Tampa Bay (+6.89\%), Atlanta (+4.31\%), and New Orleans (+4.07\%).  Anecdotally, four of the top five teams by CPAE pass defense made the playoffs in 2018 (all except for the Oakland Raiders).

\begin{figure}
    \centering
    \includegraphics[width=\textwidth]{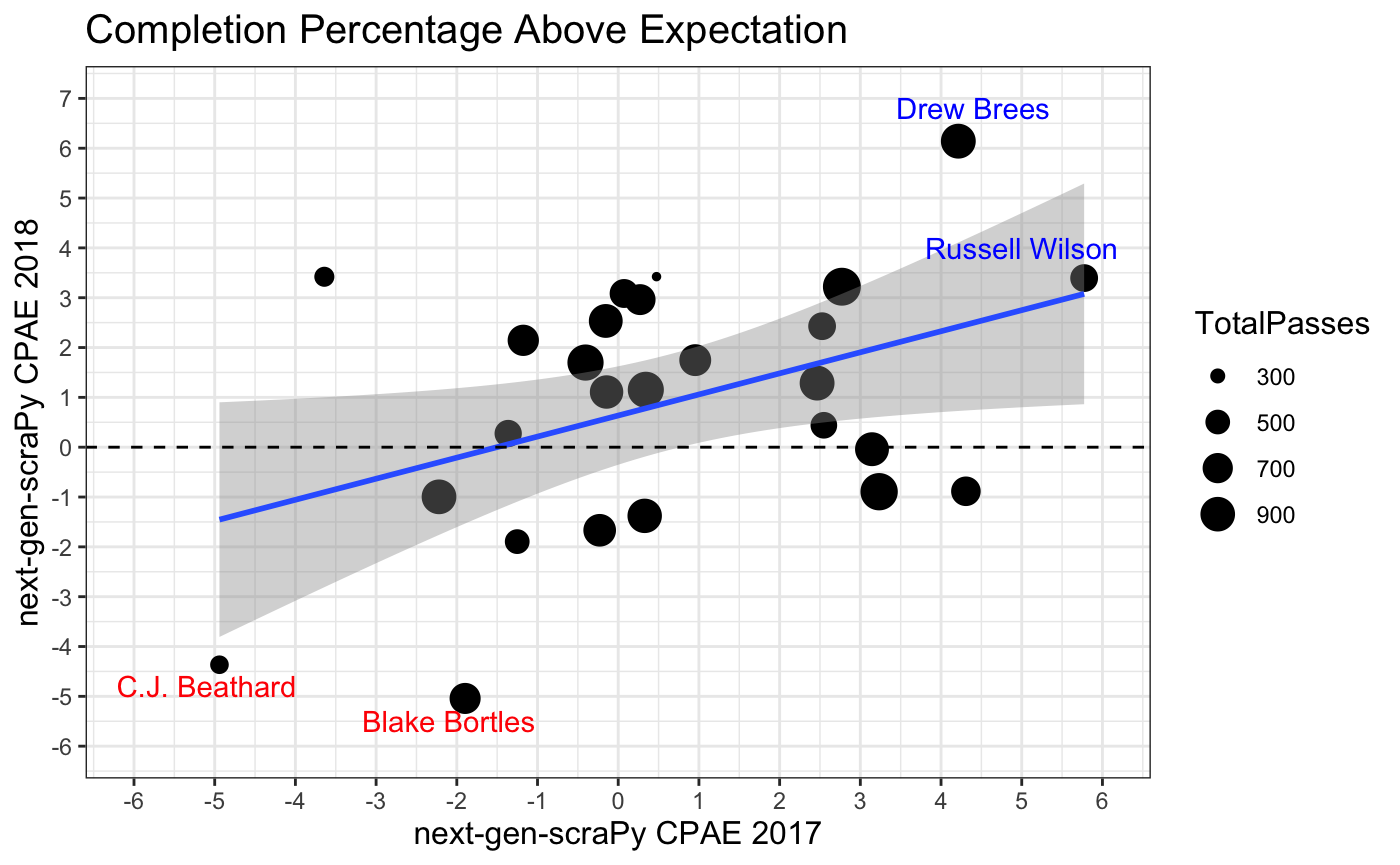}
    \caption{Stability of CPAE for QBs with at least 100 passes in 2017 and 2018.}
    \label{fig:cpae}
\end{figure}

We further explore the {\em stability} of QB $CPAE_t$ across the two seasons available in our data ($t \in \{2017, 2018\}$). 
Figure \ref{fig:cpae} presents the $CPAE$ for all qualifying QBs (i.e., at least 100 passes in each of the season). 
The linear correlation between $CPAE_{2017}$ and $CPAE_{2018}$ is 0.41 (p-val < 0.05), hence, exhibiting medium levels of stability across the seasons examined.  Of course, if we were to account for selection bias (e.g. below-average QBs tend to stay below-average, but also tend to lose their starting jobs), this relationship may actually prove stronger than stated here.

Since the NFL player tracking data became available to teams and league officials in 2016, the NFL has developed various metrics based on these data as part of their NextGen Stats (NGS) initiative. 
One of these metrics is the completion percentage above expectation\footnote{\url{ https://nextgenstats.nfl.com/stats/passing/2019/all}}, which quantifies how a QB has fared compared to the expected completion percentage of his throws. 
Contrary to our approach, the NGS CPAE takes into consideration various variables\footnote{\url{http://www.nfl.com/news/story/0ap3000000964655/article/next-gen-stats-introduction-to-completion-probability}} such as, the location of the receiver, the location of the defenders, the, distance from the sideline, etc.  NFL NGS first creates a completion probability model for each throw. 
Then, NGS CPAE is simply the difference between the actual completion result and the model's expected completion probability. 

\begin{figure}
    \centering
    \includegraphics[width = \textwidth]{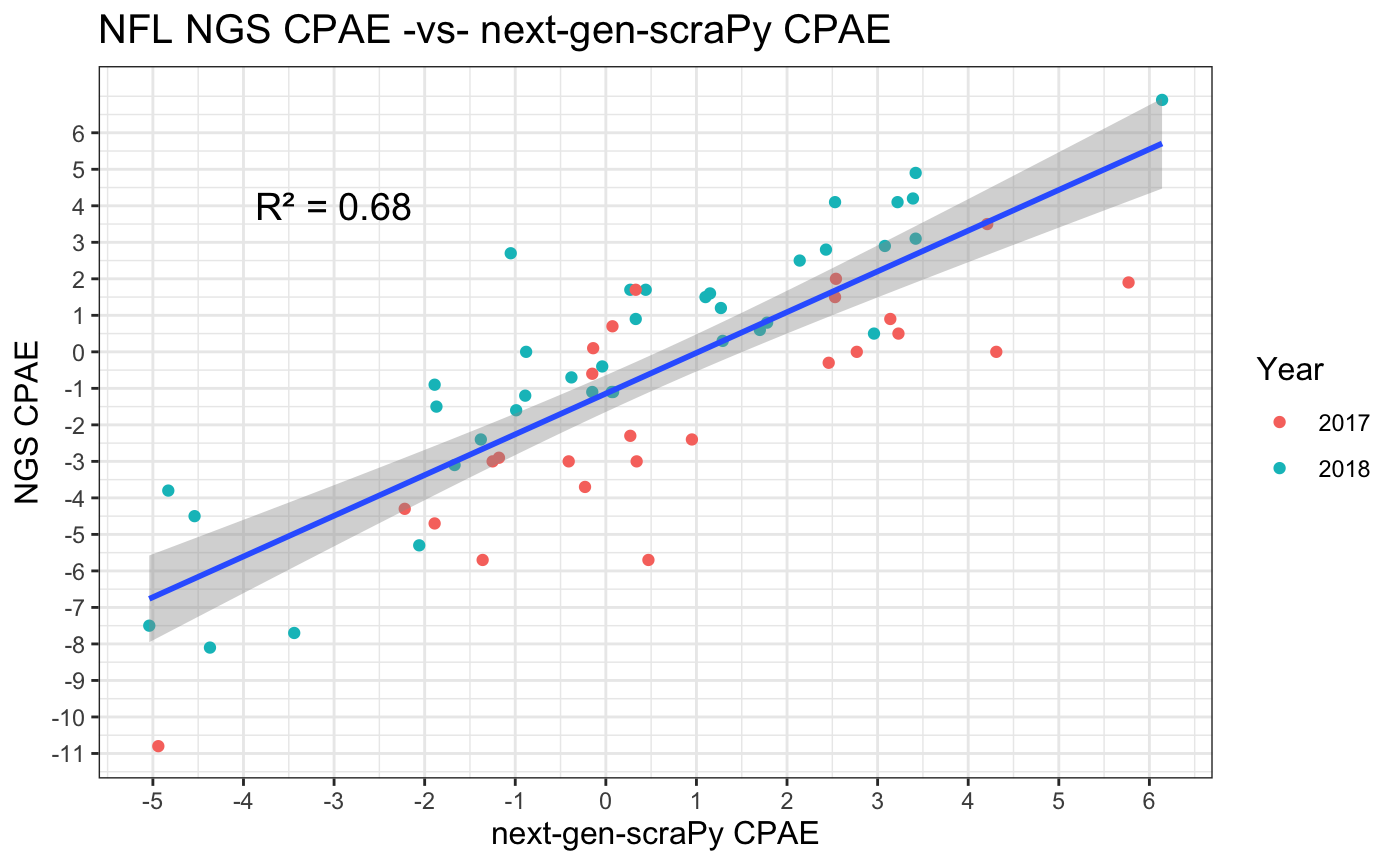}
    \caption{NFL NGS CPAE vs. {\method} CPAE in 2017 and 2018.}
    \label{fig:cpae-ngs}
\end{figure}

We collected the NGS CPAE for all QBs from the 2017 and 2018 regular seasons and compared them with the CPAE results obtained from our Naive Bayes approach. 
Figure \ref{fig:cpae-ngs} presents the results. 
Overall, the correlation between the two metrics is high  ($\rho_{2017}$= 0.81, $\rho_{2018} = 0.91$). 
Some of the differences in the CPAE values between the two approaches can be attributed to the missing passing charts that are used to create the {\method} dataset. 
Nevertheless, our version of CPAE follows closely the one provided from NGS, despite the use of only publicly available data.

Interestingly, Figure \ref{fig:cpae-ngs} also shows that NGS CPAE appears to be higher on average in 2018 than 2017, while no temporal trend exists when using {\method} CPAE.

%% file: 5_Discussion.tex
In this paper, we present {\method}, a software package that allows football fans and analysts to extract and analyze the underlying data from the pass charts provided by NFL Next Gen Stats via their player tracking technology. 
With {\method}, we implement a computer vision module that processes the images (pass charts) provided by Next Gen Stats, a $K$-Means++ clustering approach to identify passes and their locations on the field (relative to the line of scrimmage), and a DBSCAN clustering approach to remove noise from certain segmented images.  The resulting dataset contains all pass locations from the 2017 and 2018 seasons that were tracked by the NFL player and ball tracking technology and shown on their Next Gen Stats website.  This provides researchers with an abundance of data from 500+ games across two full seasons (including the most recent and relevant season), as compared to the only six weeks of data from 2017 that were temporarily made available by the NFL.

Using the resulting dataset, which we make available publicly, we build statistical models for the completion percentages by location on the field for the NFL, for individual QBs, and for team defenses.  To do this, we combine the use of generalized additive models and kernel density estimations via a two-dimensional naive Bayes approach.  Using the results of this Naive Bayes approach, we present a metric called Completion Percentage Above Expectation (CPAE).  We show that {\method} CPAE is highly correlated with NFL NGS CPAE, and appears to be more consistent across seasons.  We provide a ranking of both individual QBs and team defenses using {\method} CPAE.

Additionally, we describe a greedy record linkage approach to match passes from the {\method} dataset to passes from the NFL's tracking data, demonstrating this using the first six weeks of NFL tracking data made available via the Big Data Bowl.  Here, we find that the pass location coordinates in the {\method} closely match those from the Big Data Bowl (median difference of 1.7 yards), and we find some evidence that these deviations may potentially be attributed to one data source tracking the location of the ball, while another tracks the location of the receiver; this evidence is circumstantial, and we encourage future researchers to examine it in more detail.

While this current work only pertains to pass charts, Next Gen Stats also provides route charts for receivers, and carry charts for running backs.  In future work, we hope to extend the functionality of {\method} to extract this information and provide it to the public.  Extracting receiver routes and rusher carry paths is substantially more challenging.  We hypothesize that mixture regression models for detecting the trajectories from the image, but we have only completed very preliminary work on this topic to date.  

Next, we acknowledge that there are many alternatives to the naive Bayes approach for passing evaluation that we use here, all of which have merit and may be appropriate depending on the goal of the problem.  For example, point processes are often used in spatial statistics to model the locations of events, or the number of events in an area \citep{point}.  As such, they may be appropriate for some of the tasks in this paper (e.g. estimating pass location densities).  Similarly, given enough computing power, using regularized hierarchical generalized additive models \citep{hgam} for some tasks (e.g. estimating completion percentage surfaces for QBs or team defenses) may be more appropriate, since these can be specified with groups corresponding to individual QBs or teams and would naturally provide comparisons to average.

Finally, future researchers may improve upon our approach for linking passes in our dataset to specific plays from the NFL's tracking data.  Additionally, we encourage researchers to extend our greedy record linkage algorithm to link {\method} passes to play-by-play data from the {\tt nflscrapR} package \cite{nflscrapR}, using (for example) the air yards information provided by such data sources.  Doing so would open up many potential avenues of future research, such as estimating of spatial expected points added (EPA) or win probability added (WPA) surfaces \citep{Yurko19}. 

\pagebreak

\appendix
\section{Data Scraped from Next Gen Stats}
\label{app:appendix-A}

\begin{table}[!htb]
\begin{center}
 \begin{tabular}{p{3cm}p{10cm}}
  \toprule
 \textbf{Variable} & \textbf{Description} \\
  \midrule

completions & number of completions thrown \\

\hline
touchdowns & number of touchdowns thrown \\
\hline
attempts & number of passes thrown \\
\hline
interceptions & number of interceptions thrown \\
\hline
extraLargeImg & URL of extra-large-sized image (1200 x 1200) \\
\hline
week & week of game \\
\hline
gameId & 10-digit game identification number \\
\hline
season & NFL season \\
\hline
firstName & first name of player \\
\hline
lastName & last name of player \\
\hline
team & team name of player \\
\hline
position & position of player \\
\hline
seasonType & regular ("reg") or postseason ("post") \\
 \bottomrule
\end{tabular} 
\vspace{0.2cm}
\end{center}
\end{table}

\section{Example Subset of Data}
\label{app:appendix-B}

\begin{adjustbox}{angle=0}
\begin{tiny}
    \begin{tabular}{llllllllllll}
    \toprule
    game\_id   & team & week       & name       & pass\_type   & x\_coord     & y\_coord    & type & home\_team & away\_team & season \\
   \midrule
 2018020400 & PHI  & super-bowl & Nick Foles & COMPLETE     & -3.6  & 16.9 & post & NE         & PHI        & 2017   \\
 2018020400 & PHI  & super-bowl & Nick Foles & COMPLETE     & 16.2  & -3.0   & post & NE         & PHI        & 2017   \\
 2018020400 & PHI  & super-bowl & Nick Foles & COMPLETE     & 11.5  & -6.4 & post & NE         & PHI        & 2017   \\
 2018020400 & PHI  & super-bowl & Nick Foles & TOUCHDOWN    & -8.5  & 5.7  & post & NE         & PHI        & 2017   \\
 2018020400 & PHI  & super-bowl & Nick Foles & TOUCHDOWN    & -18.8 & 30.1 & post & NE         & PHI        & 2017   \\
 2018020400 & PHI  & super-bowl & Nick Foles & TOUCHDOWN    & -19.3 & 41.2 & post & NE         & PHI        & 2017   \\
 2018020400 & PHI  & super-bowl & Nick Foles & INTERCEPTION & 21.8  & 37.9 & post & NE         & PHI        & 2017   \\
 2018020400 & PHI  & super-bowl & Nick Foles & INCOMPLETE   & 5.1   & 7.9  & post & NE         & PHI        & 2017   \\
 2018020400 & PHI  & super-bowl & Nick Foles & INCOMPLETE   & -12.9 & 39.6 & post & NE         & PHI        & 2017   \\
 2018020400 & PHI  & super-bowl & Nick Foles & INCOMPLETE   & 26.1  & 8.0    & post & NE         & PHI  & 2017 \\
\bottomrule
\end{tabular}
\end{tiny}
\end{adjustbox} 

\section{QB CPAE}
\label{app:appendix-C}

\begin{table}[!htbp] \centering 
\begin{tabular}{@{\extracolsep{5pt}} ccccc} 
\\[-1.8ex]\hline 
\hline \\[-1.8ex] 
QB & CPAE17 & npasses\_2017 & CPAE18 & npasses\_2018 \\ 
\hline \\[-1.8ex] 
Drew Brees & 4.21 & 439 & 6.14 & 473 \\ 
Ryan Fitzpatrick & 0.47 & 112 & 3.42 & 157 \\ 
Nick Foles & -3.64 & 152 & 3.42 & 229 \\ 
Russell Wilson & 5.77 & 309 & 3.39 & 295 \\ 
Matthew Ryan & 2.77 & 524 & 3.22 & 552 \\ 
Carson Wentz & 0.07 & 333 & 3.08 & 313 \\ 
Derek Carr & 0.27 & 300 & 2.96 & 429 \\ 
Kirk Cousins & -0.15 & 394 & 2.53 & 467 \\ 
Derrick Watson & 2.53 & 110 & 2.43 & 492 \\ 
Cameron Newton & -1.18 & 352 & 2.14 & 392 \\ 
Marcus Mariota & 0.95 & 495 & 1.75 & 275 \\ 
Jared Goff & -0.41 & 428 & 1.7 & 553 \\ 
Ben Roethlisberger & 2.46 & 394 & 1.29 & 518 \\ 
Patrick Mahomes &  &  & 1.27 & 445 \\ 
Philip Rivers & 0.34 & 416 & 1.15 & 560 \\ 
Rayne Prescott & -0.14 & 408 & 1.11 & 434 \\ 
Jameis Winston & 2.55 & 268 & 0.44 & 295 \\ 
Andrew Luck &  &  & 0.33 & 559 \\ 
Mitchell Trubisky & -1.36 & 262 & 0.27 & 323 \\ 
Ryan Tannehill &  &  & 0.08 & 191 \\ 
Brock Osweiler &  &  & 0.06 & 163 \\ 
John Stafford & 3.14 & 384 & -0.04 & 480 \\ 
Aaron Rodgers &  &  & -0.15 & 573 \\ 
Baker Mayfield &  &  & -0.38 & 269 \\ 
Alexander Smith & 4.31 & 418 & -0.88 & 254 \\ 
Tom Brady & 3.23 & 524 & -0.89 & 519 \\ 
Elisha Manning & -2.22 & 369 & -1 & 536 \\ 
Sam Darnold &  &  & -1.05 & 289 \\ 
Casey Keenum & 0.33 & 382 & -1.38 & 509 \\ 
Joseph Flacco & -0.23 & 438 & -1.67 & 367 \\ 
Nicholas Mullens &  &  & -1.87 & 118 \\ 
Andrew Dalton & -1.25 & 307 & -1.89 & 195 \\ 
Lamar Jackson &  &  & -2.07 & 112 \\ 
Joshua Allen &  &  & -3.44 & 237 \\ 
Casey Beathard & -4.94 & 185 & -4.37 & 168 \\ 
Joshua Rosen &  &  & -4.54 & 260 \\ 
Jeffrey Driskel &  &  & -4.83 & 110 \\ 
Robby Bortles & -1.9 & 399 & -5.04 & 336 \\ 
\hline \\[-1.8ex] 
\caption{CPAE for 2017 and 2018 seasons for QBs with at least 100 passes in a season.} 
  \label{tab:cpae} 
\end{tabular} 
\end{table}

\section{Defense CPAE}
\label{app:appendix-D}

\begin{table}[!htbp] \centering 
\begin{tabular}{@{\extracolsep{5pt}} ccccc} 
\\[-1.8ex]\hline 
\hline \\[-1.8ex] 
Team & CPAE17 & npasses\_2017 & CPAE18 & npasses\_2018 \\ 
\hline \\[-1.8ex] 
TB & 3.54 & 380 & 6.89 & 452 \\ 
ATL & 2.36 & 524 & 4.31 & 552 \\ 
NO & -0.68 & 439 & 4.07 & 495 \\ 
DAL & 2.81 & 408 & 3.82 & 434 \\ 
IND & 1.91 & 388 & 2.94 & 559 \\ 
CIN & -2.37 & 307 & 2.52 & 305 \\ 
DET & 5.45 & 384 & 2.46 & 480 \\ 
MIN & -4.93 & 382 & 2.39 & 467 \\ 
MIA & 1.86 & 355 & 2.38 & 354 \\ 
WAS & -4.5 & 394 & 2.17 & 403 \\ 
SEA & -3.89 & 309 & 1.33 & 295 \\ 
CAR & 1.59 & 352 & 1.19 & 443 \\ 
PHI & -0.59 & 485 & 0.99 & 542 \\ 
HOU & 7 & 410 & 0.27 & 492 \\ 
ARI & -1.93 & 462 & 0.22 & 354 \\ 
SF & 0.95 & 500 & -0.33 & 344 \\ 
JAX & -3.99 & 399 & -0.7 & 401 \\ 
NE & 1.56 & 524 & -1 & 519 \\ 
GB & 7.1 & 363 & -1.06 & 573 \\ 
NYG & 1.19 & 402 & -1.08 & 536 \\ 
LAC & 0.81 & 416 & -1.11 & 560 \\ 
TEN & -0.85 & 526 & -1.13 & 323 \\ 
DEN & -1.27 & 386 & -1.13 & 509 \\ 
CLE & 3.97 & 427 & -1.27 & 339 \\ 
BUF & 2.41 & 212 & -1.48 & 327 \\ 
PIT & -1.75 & 421 & -1.65 & 526 \\ 
NYJ & -3.45 & 370 & -1.93 & 352 \\ 
KC & -1.64 & 452 & -1.98 & 445 \\ 
OAK & 1.71 & 324 & -2.12 & 429 \\ 
LA & -2.81 & 462 & -2.35 & 553 \\ 
CHI & 3.32 & 368 & -2.43 & 360 \\ 
BAL & -3.36 & 438 & -5.24 & 479 \\ 
\hline \\[-1.8ex] 
\caption{Defensive CPAE for 2017 and 2018 seasons. Lower number represents better defense.} 
  \label{tab:cpae_def} 
\end{tabular} 
\end{table}